\documentclass[aps,prl,twocolumn,superscriptaddress,showpacs]{revtex4-1}
\usepackage{graphicx}
\usepackage{dcolumn}
\usepackage{bm}
\usepackage{subfigure}
\usepackage{amsmath}
\usepackage{lipsum}
\usepackage{amssymb}
\usepackage{hyperref}
\usepackage{array}
\usepackage{times}
\usepackage{threeparttable}

\begin{document}
\title{Two-stage superconductivity in the Hatsugai-Kohmoto-BCS model}
\author{Yu Li}
\address{Kavli Institute for Theoretical Sciences, University of Chinese Academy of Sciences, Beijing 100190, China}
\author{Vivek Mishra}
\address{Kavli Institute for Theoretical Sciences, University of Chinese Academy of Sciences, Beijing 100190, China}
\author{Yi Zhou}
\address{Beijing National Laboratory for Condensed Matter Physics \& Institute of Physics, Chinese Academy of Sciences, Beijing 100190, China}
\affiliation{Songshan Lake Materials Laboratory, Dongguan, Guangdong 523808, China}
\address{CAS Center for Excellence in Topological Quantum Computation, University of Chinese Academy of Sciences, Beijing 100190, China}
\author{Fu-Chun Zhang}
\email{fuchun@ucas.ac.cn}
\address{Kavli Institute for Theoretical Sciences, University of Chinese Academy of Sciences, Beijing 100190, China}
\address{CAS Center for Excellence in Topological Quantum Computation, University of Chinese Academy of Sciences, Beijing 100190, China}

\date{\today}

\begin{abstract}
Superconductivity in strongly correlated electrons can emerge out from a normal state that is beyond the Landau's Fermi liquid paradigm, often dubbed as ``non-Fermi liquid". While the theory for non-Fermi liquid is still not yet conclusive, a recent study on the exactly-solvable Hatsugai-Kohmoto (HK) model has suggested a non-Fermi liquid ground state whose Green's function resembles the Yang-Rice-Zhang ansatz for cuprates~[P. W. Phillips, L. Yeo and E. W. Huang, Nat. Phys. {\bf 16}, 1175 (2020)].
Similar to the effect of on-site Coulomb repulsion in the Hubbard model, the repulsive interaction in the HK model divides the momentum space into three parts: empty, single-occupied and double-occupied regions, that are separated from each other by two distinct Fermi surfaces.
In the presence of an additional Bardeen-Cooper-Schrieffer (BCS)-type pairing interaction of a moderate strength, we show that the system exhibits a ``two-stage superconductivity" feature as temperature decreases: a first-order superconducting transition occurs at a temperature $T_{\rm c}$ that is followed by a sudden increase of the superconducting order parameter at a lower temperature $T_{\rm c}^{\prime}<T_{\rm c}$.
At the first stage, $T_{\rm c}^{\prime}<T<T_{\rm c}$, the pairing function arises and the entropy is released only in the vicinity of the two Fermi surfaces; while at the second stage, $T<T_{\rm c}^{\prime}$, the pairing function becomes significant and the entropy is further released in deep (single-occupied) region in the Fermi sea. The phase transitions are analyzed within the Ginzburg-Landau theory.
Our work sheds new light on unconventional superconductivity in strongly correlated electrons.

\end{abstract}

\maketitle
{\bf Introduction.} The pairing mechanism of unconventional superconductivity remains one of the central issues in condensed matter physics. Conventional superconductivity has been well captured by the classic Bardeen-Cooper-Schrieffer (BCS) theory~\cite{Bardeen1957}, in which a second-order superconducting phase transition occurs as a result of the Cooper pairing instability of the Fermi liquid normal state~\cite{Landau1959}. However, such a Fermi liquid normal state is absent in many, if not most, unconventional superconductors. Instead, the corresponding normal state is often referred as a ``non-Fermi liquid" (NFL) or ``unconventional metal" state~\cite{Schofield1999,Stewart2001,Varma2002}. In contrast to Fermi liquids that can be adiabatically connected to a gas of non-interacting fermions and be well depicted by interactions between  quasi-particles~\cite{Shankar1994,Lohneysen2007}, a generic paradigm for NFLs has not yet been established so far~\cite{Lee2018,Varma2020}. However, some experimental criteria for NFLs are commonly accepted. For instance, electric resistivity deviates from the $\rho(T)\propto{}T^2$ temperature dependence, and specific heat $C_{V}(T)$ is no longer linearly temperature-dependent~\cite{Stewart2001,Schlottmann2015,Proust2019,Greene2020}. Moreover, a variety of realistic materials exhibit NFL behaviors, which include but are not limited to cuprates~\cite{Proust2019,Greene2020}, iron-pnictides and chalcogenides~\cite{Stewart2011,Hosono2015}, and heavy-fermion compounds~\cite{Schlottmann2015,Li2021}. The superconducting phase emerges from a NFL normal state in these materials~\cite{Stewart2011,Schlottmann2015,Cooper2009}. It is illuminating to understand their pairing mechanisms from studying the pairing from a unconventional metal which beyond Landau's Fermi liquid theory.

\begin{figure}[tb]
\includegraphics[width=0.45\textwidth]{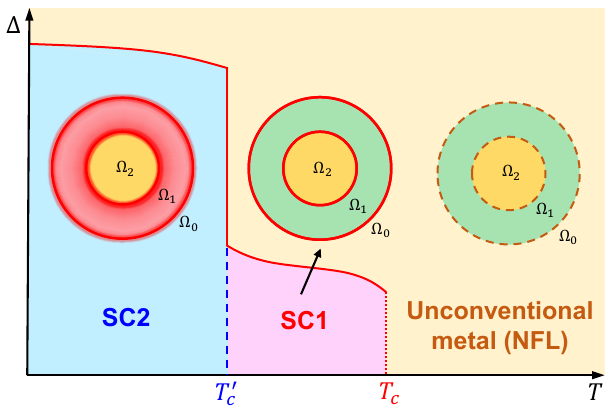}
\caption{Schematic phase diagram for the HK-BCS model. The temperature dependence of the superconducting order parameter $\Delta(T)$. (1) At $T>T_{\rm c}$, $\Delta=0$, the normal state is an unconventional metal (NFL), on which the momentum space is divided into three regions: empty ($\Omega_0$), single-occupied ($\Omega_1$) and double-occupied region ($\Omega_2$). These regions are separated by two Fermi surfaces (brown dash lines). (2) As temperature decreases, a first-order superconducting transition occurs at $T=T_{\rm c}$. For $T_{\rm c}^{\prime}<T<T_{\rm c}$, the superconductivity comes from the electron pairing in the vicinity of the two Fermi surfaces and leads to the superconducting phase in regime 1 (SC1). (3) At a lower temperature $T_{\rm c}^{\prime}$ ($T_{\rm c}^{\prime}<T_{\rm c}$), a second jump of $\Delta(T)$ takes place, resulting in the superconducting phase in regime 2 (SC2), where Cooper pairs come into being inside the singly-occupied region ($\Omega_1$) and play a significant role.
}
\label{fig1}
\end{figure}

On the theoretical side, despite the lack of a well recognized paradigm for NFLs~\cite{Coleman2001,Lee2018,Chowdhury2018,Else2021,Phillips2022}, the mechanisms and their superconducting instabilities have been extensively explored in quantum critical models from several different approaches in recent years, such as: coupling of the Fermi sea and the bosonic fluctuations~\cite{Metlitski2015,Wang2017a,Wang2017b,Damia2021}, and the system of of fermions with strong random interactions~\cite{Sachdev1993,Kitaev2015,Patel2018,Esterlis2019,Wang2020,Chowdhury2020,Inkof2022}, and the phenomenological fermion propagators with anomalous retardations~\cite{Moon2010,Wang2016,Wu2019,Abanov2020,Wu2022}, etc.. Among them, several exactly-solvable models are of particular interest that include the Hatsugai-Kohmoto (HK) model~\cite{Hatsugai1992,Lidsky1998}. The interacting part in this model can be viewed as a momentum-space counterpart to the on-site Hubbard interaction, while the non-interacting part is the same as the Hubbard model. The HK model can host a NFL state with non-Landau's quasi-particle excitations~\cite{Phillips2020,Baskaran1991}, such that it violates the Luttinger's theorem and gives rise to a Green's function that resembles the Yang-Rice-Zhang (YRZ) ansatz for cuprates~\cite{Yang2006,Rice2012}. Indeed, the zeros of the YRZ-like Green's function $G(\mathbf{k},\omega=0)$ enclose a Luttinger surface instead of a usual Fermi surface~\cite{Dzyaloshinskii2003,Konik2006,Stanescu2007,Dave2013}, indicating the Mottness in the strong-coupling limit and an unconventional metal or NFL in the region of weak or intermediate-coupling~\cite{Phillips2020,Huang2022}. The possible Cooper pairing instability and associated dynamic spectral weight transfer were also investigated~\cite{Phillips2020,Setty2020,Setty2021}.
More interestingly, it was demonstrated that Fermi arcs and a pseudo gap will show up in such an unconventional metal, when the ``on-site" interaction becomes $\bf{k}$-dependent and changes sign in momentum-space~\cite{Yang2021}.
Very recently, taking account of additional BCS pairing terms, J. Zhao {\it et al.} studied the thermodynamics of the HK-BCS model in the strong pairing limit, and revealed a first-order superconducting transition instead of the continuous phase transition in the BCS theory~\cite{Zhao2022}.

To get an intuitive picture on how the superconductivity forms in the HK-BCS model at finite temperatures, in this work, we study it in the regimes of weak and intermediate pairing strengths, which is complementary to the strong pairing limit studied in Ref.~\cite{Zhao2022}. We calculate the binding energy of a Cooper pair, and study the phase diagram. Unexpectedly, we find that the system undergoes a ``two-stage" process as temperature decreases.
As illustrated in Fig.~\ref{fig1}, in addition to a first-order superconducting transition at $T_{\rm c}$, the superconducting order parameter $\Delta(T)$ has a jump to a larger value at a lower temperature $T_{\rm c}^{\prime}\left(< T_{\rm c}\right)$, accompanying with a sudden drop in entropy. The underlying physics is interpreted in accordance with the pairing function and the entropy release in momentum space, and the nature of discontinuity in the SC order parameter and entropy as a function of temperature is analyzed in the Ginzburg-Landau theory.

\begin{figure}[tb]
\includegraphics[width=0.48\textwidth]{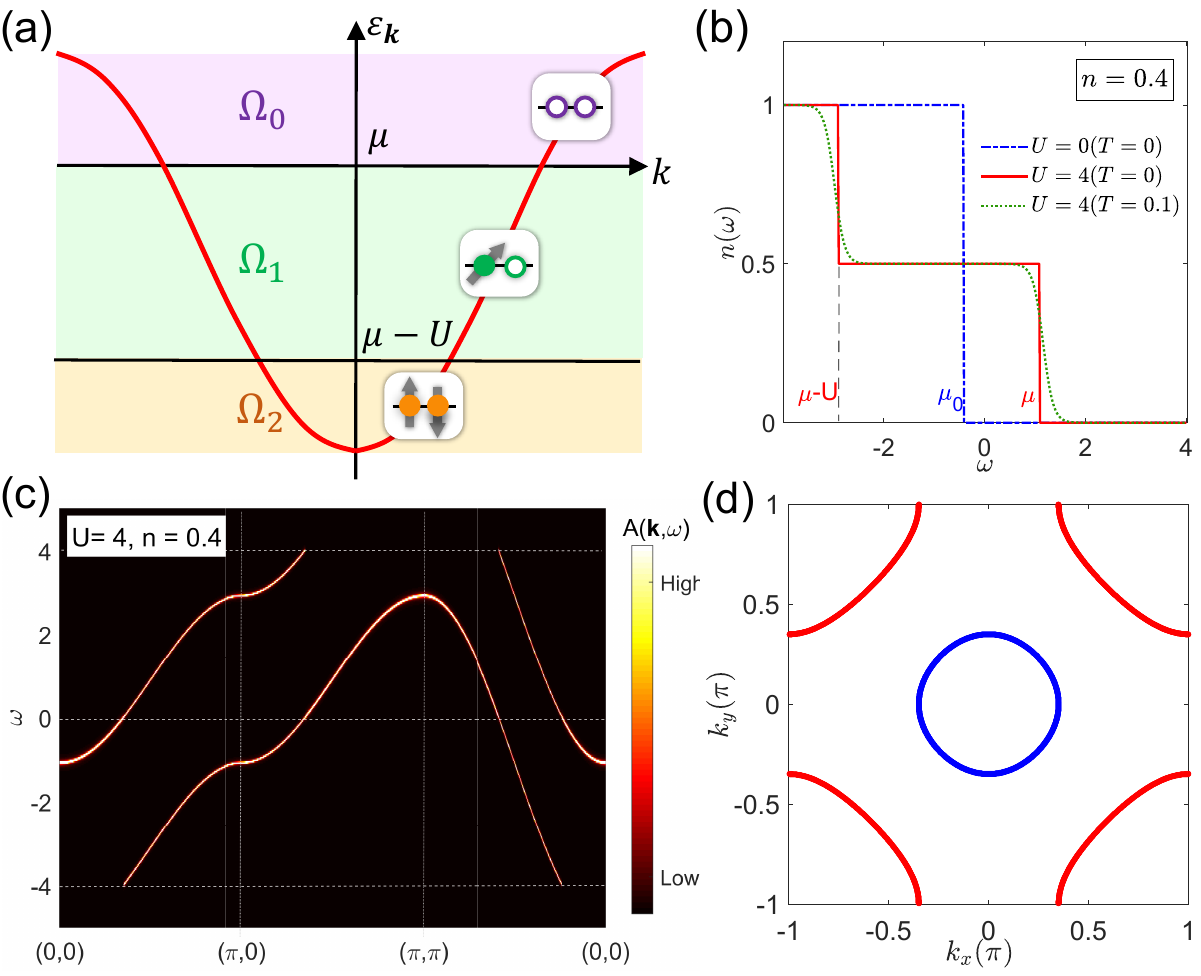}
\caption{The HK model: (a) Schematic band structure and fermion occupation in the HK model. Here $\Omega_0$, $\Omega_1$ and $\Omega_2$ represent empty, single-occupied and double-occupied regions, respectively, as sketched in Fig.~\ref{fig1}. These regions are separated by two Fermi levels at $\mu$ and $\mu-U$ respectively. (b) The electron distribution as a function of energy at filling $n=0.4$, where $\mu_0$ is the chemical potential for $U=0$. (c) The spectral function $A(\mathbf{k},\omega)$ and (d) the two Fermi surfaces for $U=4$ and $n=0.4$.}
\label{fig2}
\end{figure}

{\bf HK model revisit.} The Hatsugai-Kohmoto model~\cite{Hatsugai1992} describes strongly correlated electrons with momentum-space on-site interaction. The Hamiltonian takes a form of
\begin{equation}
H_{\text{HK}}=\sum_{\bf{k},\sigma}(\epsilon_{\bf{k}}-\mu)c^{\dagger}_{\bf{k},\sigma}c_{\bf{k},\sigma} +U\sum_{\bf{k}}n_{\bf{k}\uparrow}n_{\bf{k}\downarrow},
\end{equation}
where $c^{\dagger}_{\bf{k},\sigma}$ ($c_{\bf{k},\sigma}$) creates (annihilates) a fermion at momentum $\bf{k}$ with spin $\sigma=\uparrow,\downarrow$, and $n_{\bf{k}\sigma}=c^{\dagger}_{\bf{k},\sigma}c_{\bf{k},\sigma}$ relates to its density distribution. $\epsilon_{\bf{k}}=-2t(\cos k_x +\cos k_y)$ is the single-particle energy dispersion and $\mu$ is the chemical potential, in which $t$ is the hopping integral. Without loss of generality, we set $t=1$ as the energy unit hereafter. $U>0$ represents an on-site repulsion in the momentum space. The locality of the interaction allows us to factorize the huge Hilbert space into the direct product of the $\bf{k}$-subspace that is spanned by the basis $\{|0\rangle,c^{\dagger}_{\bf{k},\uparrow}|0\rangle,c^{\dagger}_{\bf{k},\downarrow}|0\rangle,c^{\dagger}_{\bf{k},\uparrow}c^{\dagger}_{\bf{k},\downarrow}|0\rangle\}$, making this model exactly solvable.

Ground states of the HK model can be obtained from the fermion occupation in the momentum space, as illustrated in Fig.~\ref{fig2}~(a). In the presence of a positive $U$, the momentum space will be divided into three regions in a ground state: empty region ($\Omega_0$), single-occupied region ($\Omega_1$), and double-occupied region ($\Omega_2$). This gives rise to two distinct Fermi surfaces~\cite{footnote} and two corresponding Fermi levels at $\mu$ and $\mu-U$ respectively. Here the chemical potential $\mu$ is determined by the filling number $n$ using the relation $n=\frac{1}{V_0}\sum_{\bf{k},\sigma}n_{\bf{k},\sigma}=\frac{1}{V_0}\sum_{\bf{k}}[\Theta(-\epsilon_{\bf{k}}+\mu) +\Theta(-\epsilon_{\bf{k}}+\mu-U)]$, where $\Theta$ is the Heaviside function and $V_0$ is the volume of the Brillouin zone.
The two Fermi levels can be viewed from the distribution function $n(\omega,T)\equiv\left<n_{\mathbf{k}\sigma}(\epsilon_{\mathbf{k}}=\omega,T)\right>$ as well~\cite{appendix}, where two sudden jumps occur at $\mu$ and $\mu-U$, as shown in Fig.~\ref{fig2}~(b).
When $U=0$, the region $\Omega_1$ vanishes and the two Fermi surfaces merged into a single one as in the free-fermion model. Note that the single-occupied region $\Omega_1$ always exists as long as $U>0$, while the double-occupied region $\Omega_2$ may vanish if a filling number is chosen such that $\mu-U$ exceeds the bottom of the energy band.

The retarded Green's function for this exactly solvable model reads,
\begin{equation}
G_{\sigma}\left( \mathbf{k},\omega\right) =\frac{1-\left< n_{\bf{k},\bar{\sigma}}\right>}{\omega-\xi_{\bf{k}}+i0^{+}} +\frac{\left< n_{\bf{k},\bar{\sigma}}\right>}{\omega-\xi_{\bf{k}}-U+i0^{+}},
\label{eq2}
\end{equation}

where $\xi_{\mathbf{k}}=\epsilon_{\mathbf{k}}-\mu$ and $\bar{\sigma}$ is the opposite spin index to $\sigma$. Note that $G_{\sigma}\left( \mathbf{k},\omega\right)$ does not depend on $\sigma$, and can be abbreviated as $G\left( \mathbf{k},\omega\right)$. The electron spectral function $A(\mathbf{k},\omega)=-2\text{Im}G\left( \mathbf{k},\omega\right)$ has been found at zero temperature and plotted in Fig.~\ref{fig2}~(c).
It displays two ``truncated'' bands separated by $U$, which originate from a double-occupied to single-occupied excitation and a single-occupied to empty excitation, respectively.

{\bf Residual entropy.} It is worth noting that the positive $U$ imposes the single occupancy constraint at each $\bf{k}$-point in the single-occupied region ($\Omega_1$) that gives rise to a huge ground state degeneracy and a finite entropy density at zero temperature, which is proportional to the volume of $\Omega_1$. This violates the third law of thermodynamics, and resembles the residual entropy in classical spin liquids on geometrically frustrated lattices~\cite{RMP2017}. As will be discussed later, an extra pairing interaction will lift the huge ground state degeneracy and release the entropy, resulting in a two-stage superconductivity.

\begin{figure}[tb]
\includegraphics[width=0.48\textwidth]{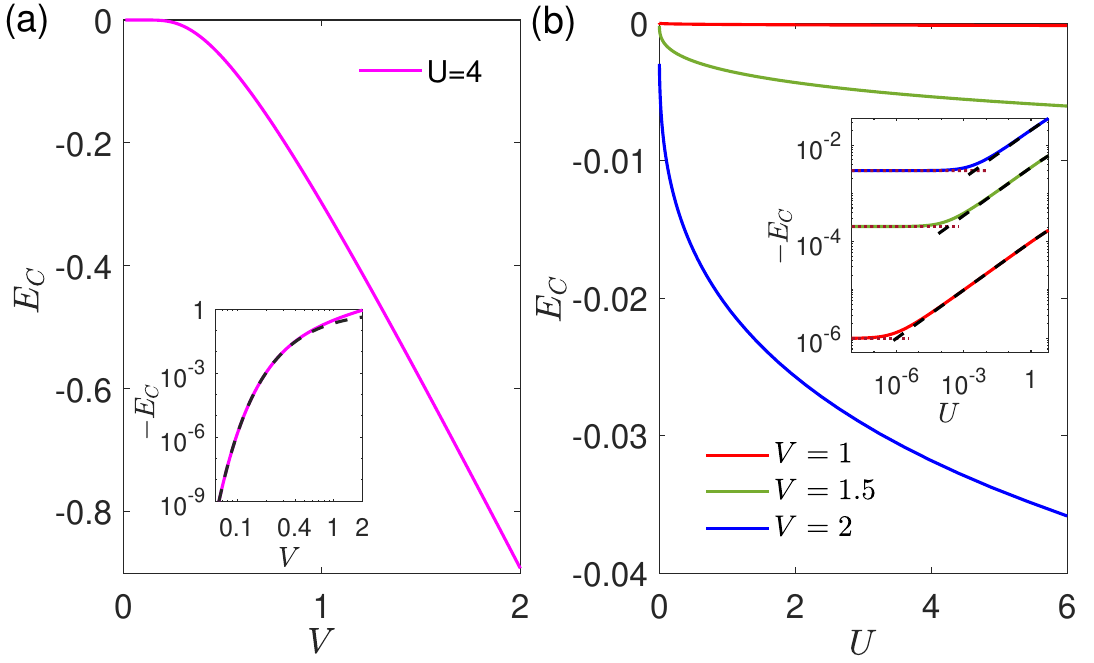}
\caption{The bound-state energy $E_{\rm C}$ for a Cooper pair in the HK model. Here the electron filling number is fixed as $n=0.4$. Solid lines represent numerical solutions to Eq.~\eqref{eq:EC0}. Black (dashed) lines show the asymptotic solution given in Eq.~\eqref{eq:EC}, where $\mu(n,U)$ is determined self-consistently. Dotted lines indicate the $U=0$ (BCS) limit. (a) $E_{\rm C}$ as a function of the pairing strength $V$, and $U=4$ is fixed. Inset: $-E_{\rm C}$ in a logarithmic plot. (b) $E_{\rm C}$ as a function of $U$. Inset: $-E_{\rm C}$ vs. $U$ in a logarithmic plot.}
\label{fig3}
\end{figure}

{\bf Cooper pair problem.} As investigated in Ref.~[\onlinecite{Phillips2020}], an infinitesimal pairing interaction will cause superconducting pairing instability in the HK model. The bound-state energy $E_{\rm C}$ for the formation of a Cooper pair on top of the Fermi sea has been estimated, where the spin polarization in the single-occupied region was assumed~\cite{Phillips2020}.
However, there is a huge spin degeneracy in the $\Omega_1$ region, and spin polarization configuration is not favorable for the Cooper pairing. Here we revisit the Cooper pair problem without assuming the spin polarization in the $\Omega_1$ region~\cite{appendix}.
Consider a generic situation when both Fermi levels locate within the bandwidth $W=8t$, thereby $-W/2<\mu<W/2$ and $U<W$, we find that the bound-state energy $E_{\rm C}$ can be determined as follows~\cite{appendix},
\begin{equation}
1=\frac{V}{4W}\ln\left\vert \frac{\left(  W-2\mu-E_{\text{C}}\right)
^{2}\left(  U-E_{\text{C}}\right)  }{E_{\text{C}} ^{3}}\right\vert. \label{eq:EC0}
\end{equation}
In the limit of $U\to{}0$, it yields $E_{\rm C}\approx-(W-2\mu)e^{-\frac{2W}{V}}$, which restores the BCS solution.
In the presence of a weak or intermediate pairing interaction $V$ and a relative large $U$, namely, when $V\ll{}W$ and $|E_{\rm C}|<U\ll{}W$, we find an asymptotic solution to Eq.~\eqref{eq:EC0},
\begin{equation}
E_{\rm C}\approx-\left(  W-2\mu{}{}\right)  ^{2/3}U^{1/3}e^{-\frac{4W}{3V}}, \label{eq:EC}
\end{equation}
which deviates from the BCS solution apparently.
For a fixed electron filling number $n$, numerical solutions to Eq.~\eqref{eq:EC0} can be found self-consistently. As plotted in Fig.~\ref{fig3},
the binding energy $|E_{\rm C}|=-E_{\rm C}$ increases as $U$ and/or $V$ increases, suggesting the enhancement of Cooper instability by the repulsive $U$. We should note that, our results are different from Ref.~\cite{Phillips2020}, in which the single-occupied region plays no role to the bind-energy, and the pairing instability is underestimated since the instability of the Fermi surface on $\mu-U$ is neglected.

{\bf HK-BCS model.} For further studying the superconductivity in the HK model, we introduce a BCS pairing interaction in the mean-field level that gives rise to the HK-BCS model as follows,
\begin{equation}
H=H_{\text{HK}}+\sum_{\bf{k}}\left(\Delta c^{\dagger}_{\bf{k}\uparrow} c^{\dagger}_{-\bf{k}\downarrow}+\text{H.c.}\right)+\frac{\Delta^2}{V},
\end{equation}
where $\Delta$ is the superconducting pairing gap and $V>0$ refers to an attractive pairing strength. This mean-field Hamiltonian can be exactly diagonalized at each $\mathbf{k}$ point, and similar to Ref.~\cite{Zhao2022}, the superconducting order parameter $\Delta\equiv-V\sum_{\mathbf{k}}\left<c_{-\mathbf{k}\downarrow}c_{\mathbf{k}\uparrow}\right>$ can be found through searching the global minimum of the free energy,
\begin{equation}
    F_{\rm S}\left[\Delta\right]=-T\ln{Z}=-T\sum_{\mathbf{k}\in\frac{1}{2}\text{BZ}}\ln{Z_{\mathbf{k}}},
\end{equation}
with the help of $\frac{\partial F_{\rm S}\left[\Delta\right]}{\partial \Delta}=0$.
Where $Z_{\mathbf{k}}\equiv\sum_{n}e^{-\beta E_{n,\mathbf{k}}}$, in which $\{\left|n,\mathbf{k}\right>\}$ and $\{E_{n,\mathbf{k}}\}$ are the eigenstates and eigenspectra obtained from diagonalizing $H$ within the tensor-product space $V_{\bf{k}}\otimes{}V_{-\bf{k}}$, where $V_{\bf{k}}$ is the subspace spanned by the basis $\{|0\rangle,c^{\dagger}_{\bf{k},\uparrow}|0\rangle,c^{\dagger}_{\bf{k},\downarrow}|0\rangle,c^{\dagger}_{\bf{k},\uparrow}c^{\dagger}_{\bf{k},\downarrow}|0\rangle\}$. In this treatment, the HK interaction remains intact and plays a crucial role in the unconventional superconductivity.

\begin{figure}[tb]
\includegraphics[width=0.46\textwidth]{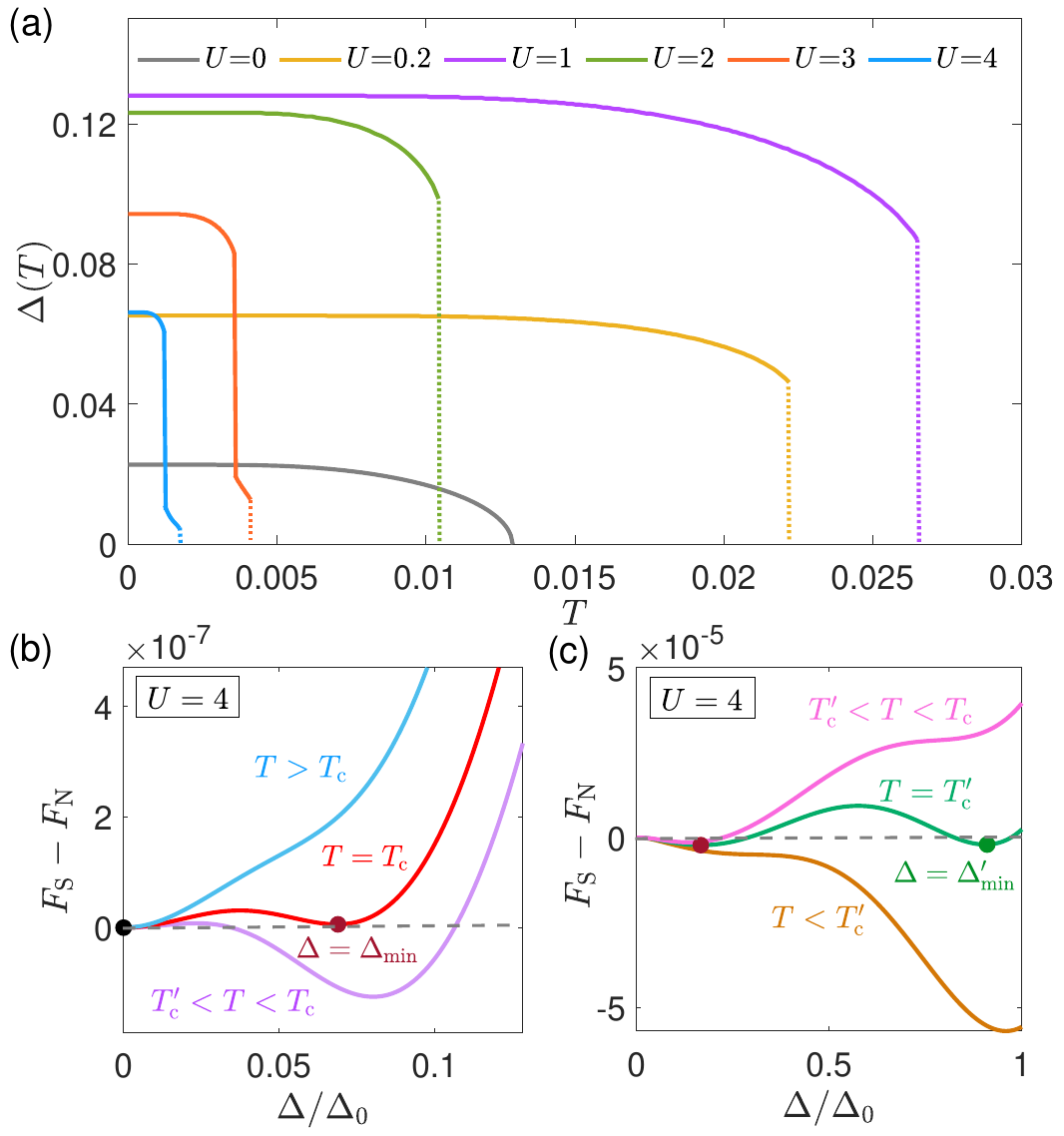}
\caption{(a) Superconducting gap as a function of temperature, where we have set $n\left(T=0\right)=0.4$ and $V=1$. Dotted lines indicate first-order transitions. For $U=4$, the free energy as a function of $\Delta$ is plotted around (b) $T=T_{\rm c}$ and (c) $T=T_{\rm c}^{\prime}$, respectively. $\Delta_0\equiv\Delta(T=0)$ and $F_{\rm N}$ is the normal state free energy that is obtained by imposing $\Delta=0$.
}
\label{fig4}
\end{figure}

{\bf Two-stage transitions.} With fixed electron filling $n$ at $T=0$ in the normal limit, and we neglect the temperature evolution of $\mu$ for simplicity, the superconducting gap $\Delta(T)$ can be found out through the minimization of the free energy for given $U$ and $V$. As observed in Ref.~[\onlinecite{Zhao2022}], there occurs a first-order superconducting phase transition as long as $U>0$, in contrast to a continuous phase transition in the $U=0$ BCS limit. Namely, as temperature is lowering, $\Delta(T)$ jumps from zero to a finite value at $T_{\rm c}$ abruptly.

Surprisingly, in addition to the first-order transition at $T_{\rm c}$, we find that there exists an extra $\Delta(T)$ jump at a lower temperature $T_{\rm c}^{\prime}\left(<T_{\rm c}\right)$ when $U$ is sufficiently large in comparison with $V$. As demonstrated in Fig.~\ref{fig4}~(a), for $n\left(T=0\right)=0.4$ and $V=1$, when $0<U<U_{\rm c}\sim{}2.7$, there is only one first-order transition at $T_{\rm c}$; while when $U>U_{\rm c}$, there emerges a sudden jump at $T_{\rm c}^{\prime}\left(<T_{\rm c}\right)$.

{\bf Free energy.} These first-order phase transitions can be understood from tracking the temperature evolution of the minimum of the free energy as a function of $\Delta$. We compute the free energy difference between the superconducting state and the normal state, $F_{\rm S}-F_{\rm N}$, and plot it in Fig.~\ref{fig4}~(b) as a function of $\Delta/\Delta_0$, where $\Delta_0=\Delta(T=0)$. Here $F_{\rm N}$ is the normal state free energy calculated at $\Delta=0$.

As shown in Figs.~\ref{fig4}(b) and \ref{fig4}(c): (1) When $T>T_{\rm c}$, the minimum of the free energy locates at $\Delta=0$; (2) When $T$ goes across $T_{\rm c}$, the free energy minimum switches from $\Delta=0$ to a finite value $\Delta=\Delta_{\text{min}}$, suggesting a first-order superconducting phase transition at $T_{\rm c}$ [see Fig.~\ref{fig4}~(a)]; (3) As temperature is lowering, in the region of $T_{\rm c}^{\prime}<T<T_{\rm c}$, there develops an extra local minimum at a larger value, $\Delta=\Delta_{\text{min}}^{\prime}\left(>\Delta_{\text{min}}\right)$, while the global minimum (i.e., the one associated with the lowest free energy) evolves from the one ($\Delta=\Delta_{\text{min}}$) arising at $T_{\rm c}^{+}$ continuously; (4) When $T$ decreases further and goes across $T_{\rm c}^{\prime}<T_{\rm c}$, the global free energy minimum switches from the smaller one $\Delta=\Delta_{\text{min}}$ to the larger one $\Delta=\Delta_{\text{min}}^{\prime}$ [see Fig.~\ref{fig4}~(c)], and then the global minimum $\Delta_{\text{min}}^{\prime}$ approaches to $\Delta=\Delta_0$ as $T\rightarrow 0$.

{\bf Phase transitions.}
Near $T_{\rm c}$, Zhao et al.~[\onlinecite{Zhao2022}] has explained the first-order nature of the SC transition via the analysis of the Ginzburg-Landau (GL) approach up to sixth-order terms of the free energy. While, in here, it is remarkable that such the two-minimum feature in free energy around $T_{\rm c}^{\prime}$ requires eighth-order terms in the GL free energy functional, which takes the form of
\begin{equation}
\delta{}\mathcal{F}\left[  \Delta\right]  =\alpha\Delta^{2}+\frac{\beta}{2}\Delta^{4}+\frac{\gamma}{3!}\Delta
^{6}+\frac{\eta}{4!}\Delta^{8}+O\left(  \Delta^{8}\right),
\end{equation}
where $\alpha$, $\beta$, $\gamma$ and $\eta$ are the expansion coefficients and depend on temperature $T$, and $\eta>0$, or $\eta=0$ and $\gamma>0$, ensures the stability of the system.
To study phase transitions for the HK-BCS model, we consider critical regions: $T\approx T_{\rm c}$ and $T\approx T_{\rm c}^{\prime}$.

(1) $T\approx{}T_{\rm c}$: It turns out that the occurrence of a first-order transition at $T=T_{\rm c}$ impose constraints for expansion coefficients at this critical point as follows~\cite{appendix},
\begin{subequations}\label{eq:Tc}
\begin{eqnarray}
\alpha &>& 0, \\
\eta & \geq& 0, \\
\frac{9\alpha\eta-2\beta\gamma}{4\gamma^2-9\beta\eta} & =&\frac{4\left(\beta^2- 2\alpha\gamma\right)}{9\alpha\eta-2\beta\gamma}>0.
\end{eqnarray}
\end{subequations}
And the superconducting gap at $T_{\rm c}$ reads
\begin{equation}
\Delta(T=T_{\rm c})=\sqrt{\frac{3\left(9\alpha\eta-2\beta\gamma\right)}{4\gamma^2-9\beta\eta}} =\sqrt{\frac{12\left(\beta^2- 2\alpha\gamma\right)}{9\alpha\eta-2\beta\gamma}}.
\end{equation}
In the limit of $\eta=0$, it becomes
\begin{equation}
\Delta(T=T_{\rm c})=\sqrt{\frac{-3\beta}{2\gamma}} =\sqrt{\frac{6\left(2\alpha\gamma-\beta^2 \right)}{\beta\gamma}},
\end{equation}
which restores the result in Ref.~\cite{Zhao2022}.

(2) $T\approx{}T_{\rm c}^{\prime}$: In the presence of the first-order-like jump at $T=T_{\rm c}^{\prime}$, the sign of expansion coefficients can be determined in the critical region as follows~\cite{appendix},
\begin{equation}
\alpha<0,~~\beta>0,~~\gamma<0~~ \text{and} ~~\eta>0.
\end{equation}
The critical condition at $T=T_{\rm c}^{\prime}$ is given by
\begin{equation}
\alpha=\gamma\left( \frac{\beta}{\eta}-\frac{\gamma^2}{3\eta^2} \right),\label{eq:Tc2}
\end{equation}
and the temperature regions $T>(<)T_{\rm c}^{\prime}$ are separated from each other in accordance with the inequality as follows,
\begin{equation}
\alpha<(>)\gamma\left( \frac{\beta}{\eta}-\frac{\gamma^2}{3\eta^2} \right), ~~ \text{for}~~ T>(<)T_{\rm c}^{\prime}.
\end{equation}

The superconducting order parameters at $T_{\rm c}^{\prime{}\pm}$ read
\begin{subequations}
\begin{align}
\Delta_{\text{min}}\equiv\Delta(T=T_{\rm c}^{\prime+})&=\sqrt{-\frac{\gamma}{\eta}-\sqrt{\frac{\gamma^2}{\eta^2}-\frac{6\alpha}{\gamma}}},\\
\Delta_{\text{min}}^{\prime}\equiv\Delta(T=T_{\rm c}^{\prime-})&=\sqrt{-\frac{\gamma}{\eta}+\sqrt{\frac{\gamma^2}{\eta^2}-\frac{6\alpha}{\gamma}}}.
\end{align}
\end{subequations}
To study the temperature dependence $\Delta(T)$ around $T_{\rm c}^{\prime}$, we introduce the dimensionless parameter $t^{\prime}=\left(T-T_{\rm c}^{\prime}\right)/T_{\rm c}^{\prime}$, and find for small $t^\prime{}$,
\begin{align}
\Delta\left(T\right)&\approx \Delta(T_{\rm c}^{\prime\pm})\left(1-b_{\pm}t^{\prime}\right)~~\text{at}~~T\gtrless{}T_{\rm c}^\prime,
\end{align}
where $b_{\pm}>0$ are two positive parameters that can be determined from experimental data or microscopic theory~\cite{appendix}.

We would like to remark that the first-order-like jump at $T=T_{\rm c}^{\prime}$ will be rounded and become a crossover when $\Delta_{\text{min}}=\Delta_{\text{min}}^{\prime}$, which leads to an extra condition for the crossover,
\begin{equation}
\gamma^3=6\alpha\eta^2, \label{eq:Tc2e}
\end{equation}
in addition to Eq.~\eqref{eq:Tc2}

\begin{figure}[tb]
\includegraphics[width=0.48\textwidth]{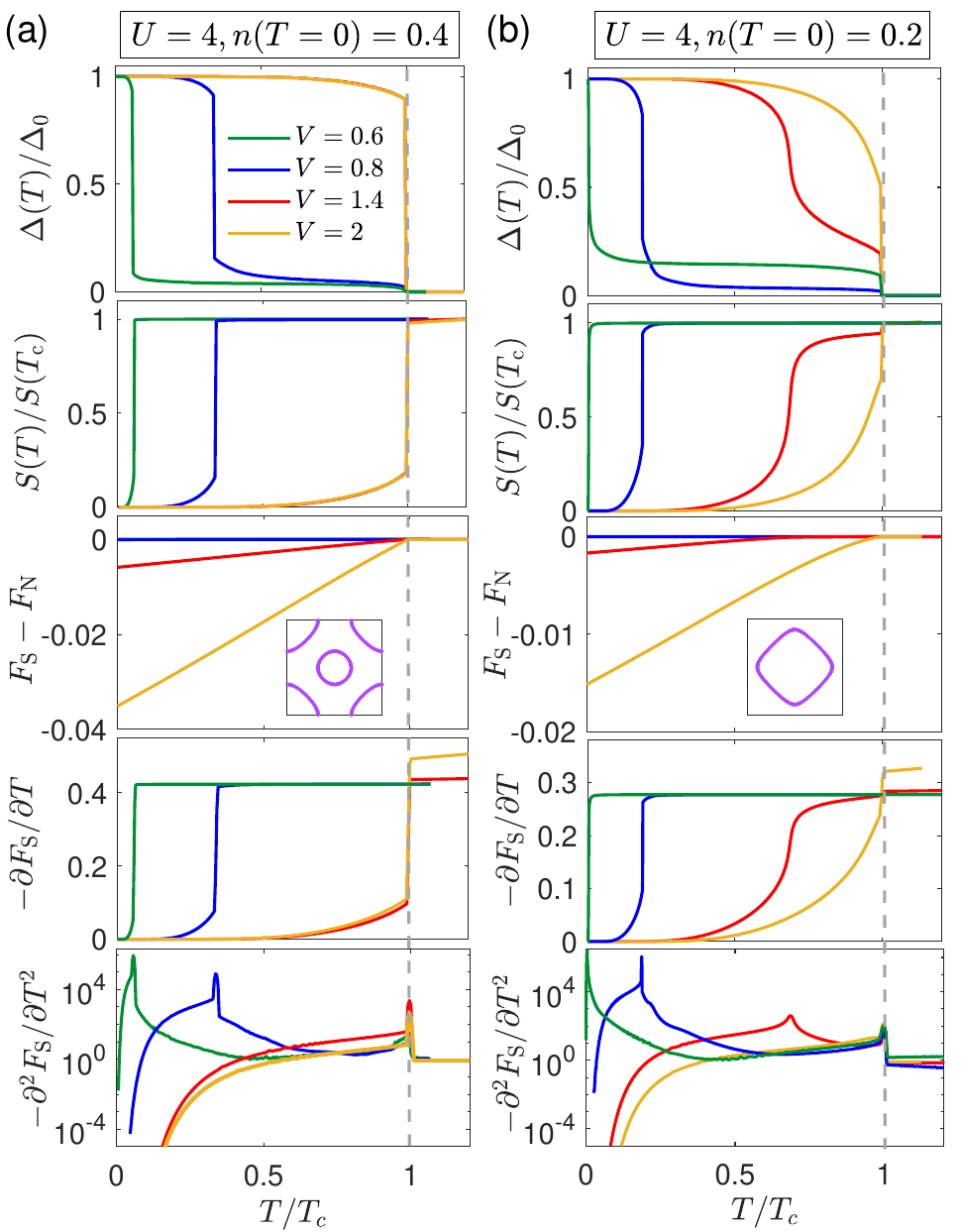}
\caption{The evolution of the relative superconducting gap $\Delta(T)/\Delta_0$, the relative entropy $S(T)/S(T_{\rm c})$, free energy difference $F_{\rm S}-F_{\rm N}$, the first-order derivative $-dF_{\rm S}/dT$ and the second-order derivative $-d^2F_{\rm S}/dT^2$ as a function of $T/T_{\rm c}$ at two fillings (a) $n\left(T=0\right)=0.4$ and (b) $n\left(T=0\right)=0.2$. Here we have set $U=4$. Insets in (a) and (b) show corresponding Fermi surfaces. For the $n\left(T=0\right)=0.2$ case, the lower Fermi level $\mu-U$ exceeds the bottom of the band, such that the double-occupied region $\Omega_2$ vanishes and there is only one Fermi surface.}
\label{fig5}
\end{figure}

{\bf Entropy release.} As mentioned earlier in this paper, the HK model has a huge residual entropy at zero temperature, which is proportional to the volume of $\Omega_1$. This residual entropy will be released by the superconducting pairing. We find that major entropy release will take place below $T_{\rm c}^{\prime}$, as long as there exist two-stage process; while there is a minor entropy release at $T_{\rm c}$. On the contrary, when there is only one first-order transition, or the two-stage process merge to a single one, there will be a significant entropy release at $T_{\rm c}$. Typical examples for entropy release have been demonstrated in Fig.~\ref{fig5}.

\begin{figure}[tb]
\includegraphics[width=0.48\textwidth]{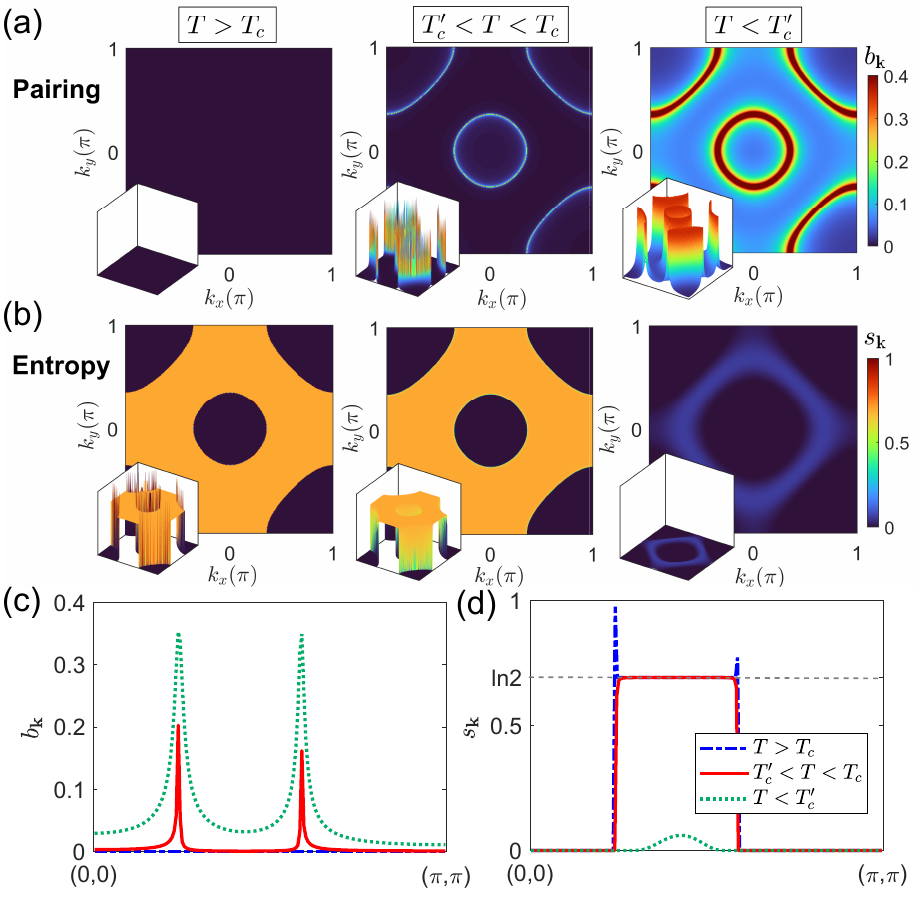}
\caption{The momentum distribution of (a) the Cooper pairing function $b_{\mathbf{k}}=\left<c_{-\mathbf{k}\downarrow}c_{\mathbf{k}\uparrow}\right>$ and (b) the entropy $s_{\mathbf{k}}=-\frac{1}{2}\sum_{n}\rho_{n,{\mathbf{k}}}\ln \rho_{n,{\mathbf{k}}}$ in the Brillouin zone.
For clear visualization, the tilted-view of (a)(b) are shown in bottom left corner of each figure. Along $(0,0)-(\pi,\pi)$, $b_{\mathbf{k}}$ and $s_{\mathbf{k}}$ are plotted in (c) and (d) respectively. Here $U=4$, $n(T=0)=0.4$ and $V=1$ have been chosen.
}
\label{fig6}
\end{figure}

{\bf Discussions.} (i) On the sudden changes at $T_{\rm c}^{\prime}$: Though the first order derivative of the free energy $\partial{}F/\partial{}T$ are discontinuous at $T_{\rm c}^{\prime}$, we didn't mark it as a real phase transition since there is only one order parameter in here. This first-order-like changes can be understood by using the Ginzburg-Landau theory as well, which can be circumvented around some critical point in the phase diagram, resembling the liquid-gas phase transition~\cite{Greiner2012}. The end of such a first-order-like change in the parameter space is indicated by the dimension reduction of the critical hyper-surface, i.e., the extra constraint in Eq.~\eqref{eq:Tc2e} reduce the dimensionality of the critical hyper-surface [given by Eq.~\eqref{eq:Tc2}] by one.

(ii) The existence of the two-stage process can be attributed to a sufficiently large $U$. Note that $U$ is the energy separation between the two Fermi levels, $\mu$ and $\mu-U$, i.e., the energy width of the single-occupied region $\Omega_1$. When $U$ is not sufficiently large, the pairing interaction $V$ will pair up all the $\mathbf{k}$ points in the small $\Omega_1$ at $T=T_{\rm c}^{-}$. Otherwise, at the fist stage, say, $T_{\rm c}^{\prime}<T<T_{\rm c}$, $V$ will pair up the states in the vicinity of Fermi surfaces only, while leave those deep inside $\Omega_1$ unpaired.

(iii) Microscopically, the two-stage superconductivity can be visualized by the momentum distribution of the Cooper pairing function $b_{\mathbf{k}}=\left<c_{-\mathbf{k}\downarrow}c_{\mathbf{k}\uparrow}\right>$ and the entropy $s_{\mathbf{k}}=-\frac{1}{2}\sum_{n}\rho_{n,{\mathbf{k}}}\ln \rho_{n,{\mathbf{k}}}$ in the Brillouin zone, where $\rho_{n\mathbf{k}}=e^{- E_{n,\mathbf{k}}/T}/Z_{\mathbf{k}}$ and the factor $\frac{1}{2}$ come from the folding of the Brillouin zone. As demonstrated in Fig.~\ref{fig6}~(a) and (b): (1) when $T>T_{\rm c}$, the pairing function $b_{\mathbf{k}}=0$, i.e., Cooper pairs are absent, and the entropy is dominated by the single-occupied region $\Omega_1$; (2) when $T_{\rm c}^{\prime}<T<T_{\rm c}$, Cooper pairs come into being in the vicinity of the two Fermi surfaces, associated with a weak and in-situ entropy release, while the entropy inside $\Omega_1$ region remains to be $\ln2$ [see Fig.~\ref{fig6}~(d)]; (3) when temperature is below $T_{\rm c}^{\prime}$, the distribution of Cooper pairs starts to extend to the whole Brillouin zone, in particular, single-occupiled $\Omega_1$ region and double occupied $\Omega_2$ region, and the residual entropy in $\Omega_1$ region are released entirely [see Fig.~\ref{fig6}~(c) and (d)].

{\bf Summary.} We have studied the HK-BCS model and revealed that, in addition to a first-order superconducting transition occurs at $T_{\rm c}$, there allows an extra first-order-like changes at a lower temperature $T_{\rm c}^{\prime}<T_{\rm c}$, as long as the momentum space on-site repulsion $U$ dominates over the superconducting pairing strength $V$. This type of two-stage process have been formulated within a Ginzburg-Landau theory consisting of eighth-order terms. The underlying physics for the formation of two-stage superconductivity has been discussed. Our results provide basic qualitative understanding of pairing in NFL systems, such as, lightly doped cuprates, heavy fermions etc. These results are also useful in understanding superconductivity in the material-specific large-scale computational methods for strongly correlated materials.

\section{Acknowledgments} The authors are grateful to S. Kirchner, F. Yang and R.-Z. Huang for helpful discussions, and we acknowledge P. W. Phillips for his useful comments. Y.Z. is supported by National Natural Science Foundation of China (No.12034004), the K. C. Wong Education Foundation (Grant No. GJTD-2020-01). Y.L. is supported by the China Postdoctoral Science Foundation (No. 2020M670422). F.C.Z is supported by National Natural Science Foundation of China (No. 11920101005).  Y.Z., V.M.\ and F.C.Z. are also supported by the Strategic Priority Research Program of Chinese Academy of Sciences (No. XDB28000000).
\section{References}

\bibliographystyle{apsrev4-1}
\bibliography{HK_BCS_ref}

\begin{thebibliography}{56}%
\makeatletter
\providecommand \@ifxundefined [1]{%
 \@ifx{#1\undefined}
}%
\providecommand \@ifnum [1]{%
 \ifnum #1\expandafter \@firstoftwo
 \else \expandafter \@secondoftwo
 \fi
}%
\providecommand \@ifx [1]{%
 \ifx #1\expandafter \@firstoftwo
 \else \expandafter \@secondoftwo
 \fi
}%
\providecommand \natexlab [1]{#1}%
\providecommand \enquote  [1]{``#1''}%
\providecommand \bibnamefont  [1]{#1}%
\providecommand \bibfnamefont [1]{#1}%
\providecommand \citenamefont [1]{#1}%
\providecommand \href@noop [0]{\@secondoftwo}%
\providecommand \href [0]{\begingroup \@sanitize@url \@href}%
\providecommand \@href[1]{\@@startlink{#1}\@@href}%
\providecommand \@@href[1]{\endgroup#1\@@endlink}%
\providecommand \@sanitize@url [0]{\catcode `\\12\catcode `\$12\catcode
  `\&12\catcode `\#12\catcode `\^12\catcode `\_12\catcode `\%12\relax}%
\providecommand \@@startlink[1]{}%
\providecommand \@@endlink[0]{}%
\providecommand \url  [0]{\begingroup\@sanitize@url \@url }%
\providecommand \@url [1]{\endgroup\@href {#1}{\urlprefix }}%
\providecommand \urlprefix  [0]{URL }%
\providecommand \Eprint [0]{\href }%
\providecommand \doibase [0]{http://dx.doi.org/}%
\providecommand \selectlanguage [0]{\@gobble}%
\providecommand \bibinfo  [0]{\@secondoftwo}%
\providecommand \bibfield  [0]{\@secondoftwo}%
\providecommand \translation [1]{[#1]}%
\providecommand \BibitemOpen [0]{}%
\providecommand \bibitemStop [0]{}%
\providecommand \bibitemNoStop [0]{.\EOS\space}%
\providecommand \EOS [0]{\spacefactor3000\relax}%
\providecommand \BibitemShut  [1]{\csname bibitem#1\endcsname}%
\let\auto@bib@innerbib\@empty
\bibitem [{\citenamefont {Bardeen}\ \emph {et~al.}(1957)\citenamefont
  {Bardeen}, \citenamefont {Cooper},\ and\ \citenamefont
  {Schrieffer}}]{Bardeen1957}%
  \BibitemOpen
  \bibfield  {author} {\bibinfo {author} {\bibfnamefont {J.}~\bibnamefont
  {Bardeen}}, \bibinfo {author} {\bibfnamefont {L.~N.}\ \bibnamefont {Cooper}},
  \ and\ \bibinfo {author} {\bibfnamefont {J.~R.}\ \bibnamefont {Schrieffer}},\
  }\href {\doibase 10.1103/PhysRev.108.1175} {\bibfield  {journal} {\bibinfo
  {journal} {Phys. Rev.}\ }\textbf {\bibinfo {volume} {108}},\ \bibinfo {pages}
  {1175} (\bibinfo {year} {1957})}\BibitemShut {NoStop}%
\bibitem [{\citenamefont {Landau}(1959)}]{Landau1959}%
  \BibitemOpen
  \bibfield  {author} {\bibinfo {author} {\bibfnamefont {L.}~\bibnamefont
  {Landau}},\ }\href {http://www.jetp.ac.ru/cgi-bin/dn/e_008_01_0070}
  {\bibfield  {journal} {\bibinfo  {journal} {Sov. Phys. JETP}\ }\textbf
  {\bibinfo {volume} {8}},\ \bibinfo {pages} {70} (\bibinfo {year}
  {1959})}\BibitemShut {NoStop}%
\bibitem [{\citenamefont {Schofield}(1999)}]{Schofield1999}%
  \BibitemOpen
  \bibfield  {author} {\bibinfo {author} {\bibfnamefont {A.~J.}\ \bibnamefont
  {Schofield}},\ }\href {https://doi.org/10.1080/001075199181602} {\bibfield
  {journal} {\bibinfo  {journal} {Contemporary Physics}\ }\textbf {\bibinfo
  {volume} {40}},\ \bibinfo {pages} {95} (\bibinfo {year} {1999})}\BibitemShut
  {NoStop}%
\bibitem [{\citenamefont {Stewart}(2001)}]{Stewart2001}%
  \BibitemOpen
  \bibfield  {author} {\bibinfo {author} {\bibfnamefont {G.~R.}\ \bibnamefont
  {Stewart}},\ }\href {\doibase 10.1103/RevModPhys.73.797} {\bibfield
  {journal} {\bibinfo  {journal} {Rev. Mod. Phys.}\ }\textbf {\bibinfo {volume}
  {73}},\ \bibinfo {pages} {797} (\bibinfo {year} {2001})}\BibitemShut
  {NoStop}%
\bibitem [{\citenamefont {Varma}\ \emph {et~al.}(2002)\citenamefont {Varma},
  \citenamefont {Nussinov},\ and\ \citenamefont {Van~Saarloos}}]{Varma2002}%
  \BibitemOpen
  \bibfield  {author} {\bibinfo {author} {\bibfnamefont {C.}~\bibnamefont
  {Varma}}, \bibinfo {author} {\bibfnamefont {Z.}~\bibnamefont {Nussinov}}, \
  and\ \bibinfo {author} {\bibfnamefont {W.}~\bibnamefont {Van~Saarloos}},\
  }\href {https://doi.org/10.1016/S0370-1573(01)00060-6} {\bibfield  {journal}
  {\bibinfo  {journal} {Physics Reports}\ }\textbf {\bibinfo {volume} {361}},\
  \bibinfo {pages} {267} (\bibinfo {year} {2002})}\BibitemShut {NoStop}%
\bibitem [{\citenamefont {Shankar}(1994)}]{Shankar1994}%
  \BibitemOpen
  \bibfield  {author} {\bibinfo {author} {\bibfnamefont {R.}~\bibnamefont
  {Shankar}},\ }\href {\doibase 10.1103/RevModPhys.66.129} {\bibfield
  {journal} {\bibinfo  {journal} {Rev. Mod. Phys.}\ }\textbf {\bibinfo {volume}
  {66}},\ \bibinfo {pages} {129} (\bibinfo {year} {1994})}\BibitemShut
  {NoStop}%
\bibitem [{\citenamefont {L\"ohneysen}\ \emph {et~al.}(2007)\citenamefont
  {L\"ohneysen}, \citenamefont {Rosch}, \citenamefont {Vojta},\ and\
  \citenamefont {W\"olfle}}]{Lohneysen2007}%
  \BibitemOpen
  \bibfield  {author} {\bibinfo {author} {\bibfnamefont {H.~v.}\ \bibnamefont
  {L\"ohneysen}}, \bibinfo {author} {\bibfnamefont {A.}~\bibnamefont {Rosch}},
  \bibinfo {author} {\bibfnamefont {M.}~\bibnamefont {Vojta}}, \ and\ \bibinfo
  {author} {\bibfnamefont {P.}~\bibnamefont {W\"olfle}},\ }\href {\doibase
  10.1103/RevModPhys.79.1015} {\bibfield  {journal} {\bibinfo  {journal} {Rev.
  Mod. Phys.}\ }\textbf {\bibinfo {volume} {79}},\ \bibinfo {pages} {1015}
  (\bibinfo {year} {2007})}\BibitemShut {NoStop}%
\bibitem [{\citenamefont {Lee}(2018)}]{Lee2018}%
  \BibitemOpen
  \bibfield  {author} {\bibinfo {author} {\bibfnamefont {S.-S.}\ \bibnamefont
  {Lee}},\ }\href {https://doi.org/10.1146/annurev-conmatphys-031016-025531}
  {\bibfield  {journal} {\bibinfo  {journal} {Annual Review of Condensed Matter
  Physics}\ }\textbf {\bibinfo {volume} {9}},\ \bibinfo {pages} {227} (\bibinfo
  {year} {2018})}\BibitemShut {NoStop}%
\bibitem [{\citenamefont {Varma}(2020)}]{Varma2020}%
  \BibitemOpen
  \bibfield  {author} {\bibinfo {author} {\bibfnamefont {C.~M.}\ \bibnamefont
  {Varma}},\ }\href {\doibase 10.1103/RevModPhys.92.031001} {\bibfield
  {journal} {\bibinfo  {journal} {Rev. Mod. Phys.}\ }\textbf {\bibinfo {volume}
  {92}},\ \bibinfo {pages} {031001} (\bibinfo {year} {2020})}\BibitemShut
  {NoStop}%
\bibitem [{\citenamefont {Schlottmann}(2015)}]{Schlottmann2015}%
  \BibitemOpen
  \bibfield  {author} {\bibinfo {author} {\bibfnamefont {P.}~\bibnamefont
  {Schlottmann}},\ }in\ \href@noop {} {\emph {\bibinfo {booktitle} {Handbook of
  Magnetic Materials}}},\ Vol.~\bibinfo {volume} {23}\ (\bibinfo  {publisher}
  {Elsevier},\ \bibinfo {year} {2015})\ pp.\ \bibinfo {pages}
  {85--160}\BibitemShut {NoStop}%
\bibitem [{\citenamefont {Proust}\ and\ \citenamefont
  {Taillefer}(2019)}]{Proust2019}%
  \BibitemOpen
  \bibfield  {author} {\bibinfo {author} {\bibfnamefont {C.}~\bibnamefont
  {Proust}}\ and\ \bibinfo {author} {\bibfnamefont {L.}~\bibnamefont
  {Taillefer}},\ }\href {\doibase 10.1146/annurev-conmatphys-031218-013210}
  {\bibfield  {journal} {\bibinfo  {journal} {Annual Review of Condensed Matter
  Physics}\ }\textbf {\bibinfo {volume} {10}},\ \bibinfo {pages} {409}
  (\bibinfo {year} {2019})}\BibitemShut {NoStop}%
\bibitem [{\citenamefont {Greene}\ \emph {et~al.}(2020)\citenamefont {Greene},
  \citenamefont {Mandal}, \citenamefont {Poniatowski},\ and\ \citenamefont
  {Sarkar}}]{Greene2020}%
  \BibitemOpen
  \bibfield  {author} {\bibinfo {author} {\bibfnamefont {R.~L.}\ \bibnamefont
  {Greene}}, \bibinfo {author} {\bibfnamefont {P.~R.}\ \bibnamefont {Mandal}},
  \bibinfo {author} {\bibfnamefont {N.~R.}\ \bibnamefont {Poniatowski}}, \ and\
  \bibinfo {author} {\bibfnamefont {T.}~\bibnamefont {Sarkar}},\ }\href
  {https://doi.org/10.1146/annurev-conmatphys-031119-050558} {\bibfield
  {journal} {\bibinfo  {journal} {Annual Review of Condensed Matter Physics}\
  }\textbf {\bibinfo {volume} {11}},\ \bibinfo {pages} {213} (\bibinfo {year}
  {2020})}\BibitemShut {NoStop}%
\bibitem [{\citenamefont {Stewart}(2011)}]{Stewart2011}%
  \BibitemOpen
  \bibfield  {author} {\bibinfo {author} {\bibfnamefont {G.~R.}\ \bibnamefont
  {Stewart}},\ }\href {\doibase 10.1103/RevModPhys.83.1589} {\bibfield
  {journal} {\bibinfo  {journal} {Rev. Mod. Phys.}\ }\textbf {\bibinfo {volume}
  {83}},\ \bibinfo {pages} {1589} (\bibinfo {year} {2011})}\BibitemShut
  {NoStop}%
\bibitem [{\citenamefont {Hosono}\ and\ \citenamefont
  {Kuroki}(2015)}]{Hosono2015}%
  \BibitemOpen
  \bibfield  {author} {\bibinfo {author} {\bibfnamefont {H.}~\bibnamefont
  {Hosono}}\ and\ \bibinfo {author} {\bibfnamefont {K.}~\bibnamefont
  {Kuroki}},\ }\href {\doibase 10.1016/j.physc.2015.02.020} {\bibfield
  {journal} {\bibinfo  {journal} {Physica C}\ }\textbf {\bibinfo {volume}
  {514}},\ \bibinfo {pages} {399} (\bibinfo {year} {2015})}\BibitemShut
  {NoStop}%
\bibitem [{\citenamefont {Li}\ \emph {et~al.}(2021)\citenamefont {Li},
  \citenamefont {Sheng},\ and\ \citenamefont {Yang}}]{Li2021}%
  \BibitemOpen
  \bibfield  {author} {\bibinfo {author} {\bibfnamefont {Y.}~\bibnamefont
  {Li}}, \bibinfo {author} {\bibfnamefont {Y.-T.}\ \bibnamefont {Sheng}}, \
  and\ \bibinfo {author} {\bibfnamefont {Y.-F.}\ \bibnamefont {Yang}},\ }\href
  {\doibase 10.7498/aps.70.20201418} {\bibfield  {journal} {\bibinfo  {journal}
  {Acta Phys. Sin.}\ }\textbf {\bibinfo {volume} {70}},\ \bibinfo {pages}
  {017402} (\bibinfo {year} {2021})}\BibitemShut {NoStop}%
\bibitem [{\citenamefont {Cooper}\ \emph {et~al.}(2009)\citenamefont {Cooper},
  \citenamefont {Wang}, \citenamefont {Vignolle}, \citenamefont {Lipscombe},
  \citenamefont {Hayden}, \citenamefont {Tanabe}, \citenamefont {Adachi},
  \citenamefont {Koike}, \citenamefont {Nohara}, \citenamefont {Takagi} \emph
  {et~al.}}]{Cooper2009}%
  \BibitemOpen
  \bibfield  {author} {\bibinfo {author} {\bibfnamefont {R.}~\bibnamefont
  {Cooper}}, \bibinfo {author} {\bibfnamefont {Y.}~\bibnamefont {Wang}},
  \bibinfo {author} {\bibfnamefont {B.}~\bibnamefont {Vignolle}}, \bibinfo
  {author} {\bibfnamefont {O.}~\bibnamefont {Lipscombe}}, \bibinfo {author}
  {\bibfnamefont {S.}~\bibnamefont {Hayden}}, \bibinfo {author} {\bibfnamefont
  {Y.}~\bibnamefont {Tanabe}}, \bibinfo {author} {\bibfnamefont
  {T.}~\bibnamefont {Adachi}}, \bibinfo {author} {\bibfnamefont
  {Y.}~\bibnamefont {Koike}}, \bibinfo {author} {\bibfnamefont
  {M.}~\bibnamefont {Nohara}}, \bibinfo {author} {\bibfnamefont
  {H.}~\bibnamefont {Takagi}},  \emph {et~al.},\ }\href
  {https://doi.org/10.1126/science.1165015} {\bibfield  {journal} {\bibinfo
  {journal} {Science}\ }\textbf {\bibinfo {volume} {323}},\ \bibinfo {pages}
  {603} (\bibinfo {year} {2009})}\BibitemShut {NoStop}%
\bibitem [{\citenamefont {Coleman}\ \emph {et~al.}(2001)\citenamefont
  {Coleman}, \citenamefont {P{\'e}pin}, \citenamefont {Si},\ and\ \citenamefont
  {Ramazashvili}}]{Coleman2001}%
  \BibitemOpen
  \bibfield  {author} {\bibinfo {author} {\bibfnamefont {P.}~\bibnamefont
  {Coleman}}, \bibinfo {author} {\bibfnamefont {C.}~\bibnamefont {P{\'e}pin}},
  \bibinfo {author} {\bibfnamefont {Q.}~\bibnamefont {Si}}, \ and\ \bibinfo
  {author} {\bibfnamefont {R.}~\bibnamefont {Ramazashvili}},\ }\href
  {https://iopscience.iop.org/article/10.1088/0953-8984/13/35/202/meta}
  {\bibfield  {journal} {\bibinfo  {journal} {Journal of Physics: Condensed
  Matter}\ }\textbf {\bibinfo {volume} {13}},\ \bibinfo {pages} {R723}
  (\bibinfo {year} {2001})}\BibitemShut {NoStop}%
\bibitem [{\citenamefont {Chowdhury}\ \emph {et~al.}(2018)\citenamefont
  {Chowdhury}, \citenamefont {Werman}, \citenamefont {Berg},\ and\
  \citenamefont {Senthil}}]{Chowdhury2018}%
  \BibitemOpen
  \bibfield  {author} {\bibinfo {author} {\bibfnamefont {D.}~\bibnamefont
  {Chowdhury}}, \bibinfo {author} {\bibfnamefont {Y.}~\bibnamefont {Werman}},
  \bibinfo {author} {\bibfnamefont {E.}~\bibnamefont {Berg}}, \ and\ \bibinfo
  {author} {\bibfnamefont {T.}~\bibnamefont {Senthil}},\ }\href {\doibase
  10.1103/PhysRevX.8.031024} {\bibfield  {journal} {\bibinfo  {journal} {Phys.
  Rev. X}\ }\textbf {\bibinfo {volume} {8}},\ \bibinfo {pages} {031024}
  (\bibinfo {year} {2018})}\BibitemShut {NoStop}%
\bibitem [{\citenamefont {Else}\ \emph {et~al.}(2021)\citenamefont {Else},
  \citenamefont {Thorngren},\ and\ \citenamefont {Senthil}}]{Else2021}%
  \BibitemOpen
  \bibfield  {author} {\bibinfo {author} {\bibfnamefont {D.~V.}\ \bibnamefont
  {Else}}, \bibinfo {author} {\bibfnamefont {R.}~\bibnamefont {Thorngren}}, \
  and\ \bibinfo {author} {\bibfnamefont {T.}~\bibnamefont {Senthil}},\ }\href
  {\doibase 10.1103/PhysRevX.11.021005} {\bibfield  {journal} {\bibinfo
  {journal} {Phys. Rev. X}\ }\textbf {\bibinfo {volume} {11}},\ \bibinfo
  {pages} {021005} (\bibinfo {year} {2021})}\BibitemShut {NoStop}%
\bibitem [{\citenamefont {Phillips}\ \emph {et~al.}(2022)\citenamefont
  {Phillips}, \citenamefont {Hussey},\ and\ \citenamefont
  {Abbamonte}}]{Phillips2022}%
  \BibitemOpen
  \bibfield  {author} {\bibinfo {author} {\bibfnamefont {P.~W.}\ \bibnamefont
  {Phillips}}, \bibinfo {author} {\bibfnamefont {N.~E.}\ \bibnamefont
  {Hussey}}, \ and\ \bibinfo {author} {\bibfnamefont {P.}~\bibnamefont
  {Abbamonte}},\ }\href {\doibase 10.1126/science.abh4273} {\bibfield
  {journal} {\bibinfo  {journal} {Science}\ }\textbf {\bibinfo {volume}
  {377}},\ \bibinfo {pages} {eabh4273} (\bibinfo {year} {2022})}\BibitemShut
  {NoStop}%
\bibitem [{\citenamefont {Metlitski}\ \emph {et~al.}(2015)\citenamefont
  {Metlitski}, \citenamefont {Mross}, \citenamefont {Sachdev},\ and\
  \citenamefont {Senthil}}]{Metlitski2015}%
  \BibitemOpen
  \bibfield  {author} {\bibinfo {author} {\bibfnamefont {M.~A.}\ \bibnamefont
  {Metlitski}}, \bibinfo {author} {\bibfnamefont {D.~F.}\ \bibnamefont
  {Mross}}, \bibinfo {author} {\bibfnamefont {S.}~\bibnamefont {Sachdev}}, \
  and\ \bibinfo {author} {\bibfnamefont {T.}~\bibnamefont {Senthil}},\ }\href
  {\doibase 10.1103/PhysRevB.91.115111} {\bibfield  {journal} {\bibinfo
  {journal} {Phys. Rev. B}\ }\textbf {\bibinfo {volume} {91}},\ \bibinfo
  {pages} {115111} (\bibinfo {year} {2015})}\BibitemShut {NoStop}%
\bibitem [{\citenamefont {Wang}\ \emph
  {et~al.}(2017{\natexlab{a}})\citenamefont {Wang}, \citenamefont {Schattner},
  \citenamefont {Berg},\ and\ \citenamefont {Fernandes}}]{Wang2017a}%
  \BibitemOpen
  \bibfield  {author} {\bibinfo {author} {\bibfnamefont {X.}~\bibnamefont
  {Wang}}, \bibinfo {author} {\bibfnamefont {Y.}~\bibnamefont {Schattner}},
  \bibinfo {author} {\bibfnamefont {E.}~\bibnamefont {Berg}}, \ and\ \bibinfo
  {author} {\bibfnamefont {R.~M.}\ \bibnamefont {Fernandes}},\ }\href {\doibase
  10.1103/PhysRevB.95.174520} {\bibfield  {journal} {\bibinfo  {journal} {Phys.
  Rev. B}\ }\textbf {\bibinfo {volume} {95}},\ \bibinfo {pages} {174520}
  (\bibinfo {year} {2017}{\natexlab{a}})}\BibitemShut {NoStop}%
\bibitem [{\citenamefont {Wang}\ \emph
  {et~al.}(2017{\natexlab{b}})\citenamefont {Wang}, \citenamefont {Raghu},\
  and\ \citenamefont {Torroba}}]{Wang2017b}%
  \BibitemOpen
  \bibfield  {author} {\bibinfo {author} {\bibfnamefont {H.}~\bibnamefont
  {Wang}}, \bibinfo {author} {\bibfnamefont {S.}~\bibnamefont {Raghu}}, \ and\
  \bibinfo {author} {\bibfnamefont {G.}~\bibnamefont {Torroba}},\ }\href
  {\doibase 10.1103/PhysRevB.95.165137} {\bibfield  {journal} {\bibinfo
  {journal} {Phys. Rev. B}\ }\textbf {\bibinfo {volume} {95}},\ \bibinfo
  {pages} {165137} (\bibinfo {year} {2017}{\natexlab{b}})}\BibitemShut
  {NoStop}%
\bibitem [{\citenamefont {Damia}\ \emph {et~al.}(2021)\citenamefont {Damia},
  \citenamefont {Sol\'{\i}s},\ and\ \citenamefont {Torroba}}]{Damia2021}%
  \BibitemOpen
  \bibfield  {author} {\bibinfo {author} {\bibfnamefont {J.~A.}\ \bibnamefont
  {Damia}}, \bibinfo {author} {\bibfnamefont {M.}~\bibnamefont {Sol\'{\i}s}}, \
  and\ \bibinfo {author} {\bibfnamefont {G.}~\bibnamefont {Torroba}},\ }\href
  {\doibase 10.1103/PhysRevB.103.155161} {\bibfield  {journal} {\bibinfo
  {journal} {Phys. Rev. B}\ }\textbf {\bibinfo {volume} {103}},\ \bibinfo
  {pages} {155161} (\bibinfo {year} {2021})}\BibitemShut {NoStop}%
\bibitem [{\citenamefont {Sachdev}\ and\ \citenamefont
  {Ye}(1993)}]{Sachdev1993}%
  \BibitemOpen
  \bibfield  {author} {\bibinfo {author} {\bibfnamefont {S.}~\bibnamefont
  {Sachdev}}\ and\ \bibinfo {author} {\bibfnamefont {J.}~\bibnamefont {Ye}},\
  }\href {\doibase 10.1103/PhysRevLett.70.3339} {\bibfield  {journal} {\bibinfo
   {journal} {Phys. Rev. Lett.}\ }\textbf {\bibinfo {volume} {70}},\ \bibinfo
  {pages} {3339} (\bibinfo {year} {1993})}\BibitemShut {NoStop}%
\bibitem [{\citenamefont {Kitaev}(2015)}]{Kitaev2015}%
  \BibitemOpen
  \bibfield  {author} {\bibinfo {author} {\bibfnamefont {A.}~\bibnamefont
  {Kitaev}},\ }\href {http://online.kitp.ucsb.edu/online/entangled15/kitaev/}
  {\bibfield  {journal} {\bibinfo  {journal} {Talk at KITP, University of
  California, Santa Barbara}\ } (\bibinfo {year} {2015})}\BibitemShut {NoStop}%
\bibitem [{\citenamefont {Patel}\ \emph {et~al.}(2018)\citenamefont {Patel},
  \citenamefont {Lawler},\ and\ \citenamefont {Kim}}]{Patel2018}%
  \BibitemOpen
  \bibfield  {author} {\bibinfo {author} {\bibfnamefont {A.~A.}\ \bibnamefont
  {Patel}}, \bibinfo {author} {\bibfnamefont {M.~J.}\ \bibnamefont {Lawler}}, \
  and\ \bibinfo {author} {\bibfnamefont {E.-A.}\ \bibnamefont {Kim}},\ }\href
  {\doibase 10.1103/PhysRevLett.121.187001} {\bibfield  {journal} {\bibinfo
  {journal} {Phys. Rev. Lett.}\ }\textbf {\bibinfo {volume} {121}},\ \bibinfo
  {pages} {187001} (\bibinfo {year} {2018})}\BibitemShut {NoStop}%
\bibitem [{\citenamefont {Esterlis}\ and\ \citenamefont
  {Schmalian}(2019)}]{Esterlis2019}%
  \BibitemOpen
  \bibfield  {author} {\bibinfo {author} {\bibfnamefont {I.}~\bibnamefont
  {Esterlis}}\ and\ \bibinfo {author} {\bibfnamefont {J.}~\bibnamefont
  {Schmalian}},\ }\href {\doibase 10.1103/PhysRevB.100.115132} {\bibfield
  {journal} {\bibinfo  {journal} {Phys. Rev. B}\ }\textbf {\bibinfo {volume}
  {100}},\ \bibinfo {pages} {115132} (\bibinfo {year} {2019})}\BibitemShut
  {NoStop}%
\bibitem [{\citenamefont {Wang}(2020)}]{Wang2020}%
  \BibitemOpen
  \bibfield  {author} {\bibinfo {author} {\bibfnamefont {Y.}~\bibnamefont
  {Wang}},\ }\href {\doibase 10.1103/PhysRevLett.124.017002} {\bibfield
  {journal} {\bibinfo  {journal} {Phys. Rev. Lett.}\ }\textbf {\bibinfo
  {volume} {124}},\ \bibinfo {pages} {017002} (\bibinfo {year}
  {2020})}\BibitemShut {NoStop}%
\bibitem [{\citenamefont {Chowdhury}\ and\ \citenamefont
  {Berg}(2020)}]{Chowdhury2020}%
  \BibitemOpen
  \bibfield  {author} {\bibinfo {author} {\bibfnamefont {D.}~\bibnamefont
  {Chowdhury}}\ and\ \bibinfo {author} {\bibfnamefont {E.}~\bibnamefont
  {Berg}},\ }\href {\doibase 10.1103/PhysRevResearch.2.013301} {\bibfield
  {journal} {\bibinfo  {journal} {Phys. Rev. Research}\ }\textbf {\bibinfo
  {volume} {2}},\ \bibinfo {pages} {013301} (\bibinfo {year}
  {2020})}\BibitemShut {NoStop}%
\bibitem [{\citenamefont {Inkof}\ \emph {et~al.}(2022)\citenamefont {Inkof},
  \citenamefont {Schalm},\ and\ \citenamefont {Schmalian}}]{Inkof2022}%
  \BibitemOpen
  \bibfield  {author} {\bibinfo {author} {\bibfnamefont {G.-A.}\ \bibnamefont
  {Inkof}}, \bibinfo {author} {\bibfnamefont {K.}~\bibnamefont {Schalm}}, \
  and\ \bibinfo {author} {\bibfnamefont {J.}~\bibnamefont {Schmalian}},\ }\href
  {\doibase 10.1038/s41535-022-00460-8} {\bibfield  {journal} {\bibinfo
  {journal} {npj Quantum Materials}\ }\textbf {\bibinfo {volume} {7}},\
  \bibinfo {pages} {1} (\bibinfo {year} {2022})}\BibitemShut {NoStop}%
\bibitem [{\citenamefont {Moon}\ and\ \citenamefont
  {Chubukov}(2010)}]{Moon2010}%
  \BibitemOpen
  \bibfield  {author} {\bibinfo {author} {\bibfnamefont {E.-G.}\ \bibnamefont
  {Moon}}\ and\ \bibinfo {author} {\bibfnamefont {A.}~\bibnamefont
  {Chubukov}},\ }\href {\doibase 10.1007/s10909-010-0199-y} {\bibfield
  {journal} {\bibinfo  {journal} {Journal of Low Temperature Physics}\ }\textbf
  {\bibinfo {volume} {161}},\ \bibinfo {pages} {263} (\bibinfo {year}
  {2010})}\BibitemShut {NoStop}%
\bibitem [{\citenamefont {Wang}\ \emph {et~al.}(2016)\citenamefont {Wang},
  \citenamefont {Abanov}, \citenamefont {Altshuler}, \citenamefont
  {Yuzbashyan},\ and\ \citenamefont {Chubukov}}]{Wang2016}%
  \BibitemOpen
  \bibfield  {author} {\bibinfo {author} {\bibfnamefont {Y.}~\bibnamefont
  {Wang}}, \bibinfo {author} {\bibfnamefont {A.}~\bibnamefont {Abanov}},
  \bibinfo {author} {\bibfnamefont {B.~L.}\ \bibnamefont {Altshuler}}, \bibinfo
  {author} {\bibfnamefont {E.~A.}\ \bibnamefont {Yuzbashyan}}, \ and\ \bibinfo
  {author} {\bibfnamefont {A.~V.}\ \bibnamefont {Chubukov}},\ }\href {\doibase
  10.1103/PhysRevLett.117.157001} {\bibfield  {journal} {\bibinfo  {journal}
  {Phys. Rev. Lett.}\ }\textbf {\bibinfo {volume} {117}},\ \bibinfo {pages}
  {157001} (\bibinfo {year} {2016})}\BibitemShut {NoStop}%
\bibitem [{\citenamefont {Wu}\ \emph {et~al.}(2019)\citenamefont {Wu},
  \citenamefont {Abanov}, \citenamefont {Wang},\ and\ \citenamefont
  {Chubukov}}]{Wu2019}%
  \BibitemOpen
  \bibfield  {author} {\bibinfo {author} {\bibfnamefont {Y.-M.}\ \bibnamefont
  {Wu}}, \bibinfo {author} {\bibfnamefont {A.}~\bibnamefont {Abanov}}, \bibinfo
  {author} {\bibfnamefont {Y.}~\bibnamefont {Wang}}, \ and\ \bibinfo {author}
  {\bibfnamefont {A.~V.}\ \bibnamefont {Chubukov}},\ }\href {\doibase
  10.1103/PhysRevB.99.144512} {\bibfield  {journal} {\bibinfo  {journal} {Phys.
  Rev. B}\ }\textbf {\bibinfo {volume} {99}},\ \bibinfo {pages} {144512}
  (\bibinfo {year} {2019})}\BibitemShut {NoStop}%
\bibitem [{\citenamefont {Abanov}\ and\ \citenamefont
  {Chubukov}(2020)}]{Abanov2020}%
  \BibitemOpen
  \bibfield  {author} {\bibinfo {author} {\bibfnamefont {A.}~\bibnamefont
  {Abanov}}\ and\ \bibinfo {author} {\bibfnamefont {A.~V.}\ \bibnamefont
  {Chubukov}},\ }\href {\doibase 10.1103/PhysRevB.102.024524} {\bibfield
  {journal} {\bibinfo  {journal} {Phys. Rev. B}\ }\textbf {\bibinfo {volume}
  {102}},\ \bibinfo {pages} {024524} (\bibinfo {year} {2020})}\BibitemShut
  {NoStop}%
\bibitem [{\citenamefont {Wu}\ \emph {et~al.}(2022)\citenamefont {Wu},
  \citenamefont {Zhang}, \citenamefont {Abanov},\ and\ \citenamefont
  {Chubukov}}]{Wu2022}%
  \BibitemOpen
  \bibfield  {author} {\bibinfo {author} {\bibfnamefont {Y.-M.}\ \bibnamefont
  {Wu}}, \bibinfo {author} {\bibfnamefont {S.-S.}\ \bibnamefont {Zhang}},
  \bibinfo {author} {\bibfnamefont {A.}~\bibnamefont {Abanov}}, \ and\ \bibinfo
  {author} {\bibfnamefont {A.~V.}\ \bibnamefont {Chubukov}},\ }\href@noop {}
  {\bibfield  {journal} {\bibinfo  {journal} {arXiv preprint arXiv:2205.10903}\
  } (\bibinfo {year} {2022})}\BibitemShut {NoStop}%
\bibitem [{\citenamefont {Hatsugai}\ and\ \citenamefont
  {Kohmoto}(1992)}]{Hatsugai1992}%
  \BibitemOpen
  \bibfield  {author} {\bibinfo {author} {\bibfnamefont {Y.}~\bibnamefont
  {Hatsugai}}\ and\ \bibinfo {author} {\bibfnamefont {M.}~\bibnamefont
  {Kohmoto}},\ }\href {\doibase 10.1143/JPSJ.61.2056} {\bibfield  {journal}
  {\bibinfo  {journal} {Journal of the Physical Society of Japan}\ }\textbf
  {\bibinfo {volume} {61}},\ \bibinfo {pages} {2056} (\bibinfo {year}
  {1992})}\BibitemShut {NoStop}%
\bibitem [{\citenamefont {Lidsky}\ \emph {et~al.}(1998)\citenamefont {Lidsky},
  \citenamefont {Shiraishi}, \citenamefont {Hatsugai},\ and\ \citenamefont
  {Kohmoto}}]{Lidsky1998}%
  \BibitemOpen
  \bibfield  {author} {\bibinfo {author} {\bibfnamefont {D.}~\bibnamefont
  {Lidsky}}, \bibinfo {author} {\bibfnamefont {J.}~\bibnamefont {Shiraishi}},
  \bibinfo {author} {\bibfnamefont {Y.}~\bibnamefont {Hatsugai}}, \ and\
  \bibinfo {author} {\bibfnamefont {M.}~\bibnamefont {Kohmoto}},\ }\href
  {\doibase 10.1103/PhysRevB.57.1340} {\bibfield  {journal} {\bibinfo
  {journal} {Phys. Rev. B}\ }\textbf {\bibinfo {volume} {57}},\ \bibinfo
  {pages} {1340} (\bibinfo {year} {1998})}\BibitemShut {NoStop}%
\bibitem [{\citenamefont {Phillips}\ \emph {et~al.}(2020)\citenamefont
  {Phillips}, \citenamefont {Yeo},\ and\ \citenamefont {Huang}}]{Phillips2020}%
  \BibitemOpen
  \bibfield  {author} {\bibinfo {author} {\bibfnamefont {P.~W.}\ \bibnamefont
  {Phillips}}, \bibinfo {author} {\bibfnamefont {L.}~\bibnamefont {Yeo}}, \
  and\ \bibinfo {author} {\bibfnamefont {E.~W.}\ \bibnamefont {Huang}},\ }\href
  {\doibase 10.1038/s41567-020-0988-4} {\bibfield  {journal} {\bibinfo
  {journal} {Nature Physics}\ }\textbf {\bibinfo {volume} {16}},\ \bibinfo
  {pages} {1175} (\bibinfo {year} {2020})}\BibitemShut {NoStop}%
\bibitem [{\citenamefont {Baskaran}(1991)}]{Baskaran1991}%
  \BibitemOpen
  \bibfield  {author} {\bibinfo {author} {\bibfnamefont {G.}~\bibnamefont
  {Baskaran}},\ }\href {https://doi.org/10.1142/S0217984991000782} {\bibfield
  {journal} {\bibinfo  {journal} {Modern Physics Letters B}\ }\textbf {\bibinfo
  {volume} {5}},\ \bibinfo {pages} {643} (\bibinfo {year} {1991})}\BibitemShut
  {NoStop}%
\bibitem [{\citenamefont {Yang}\ \emph {et~al.}(2006)\citenamefont {Yang},
  \citenamefont {Rice},\ and\ \citenamefont {Zhang}}]{Yang2006}%
  \BibitemOpen
  \bibfield  {author} {\bibinfo {author} {\bibfnamefont {K.-Y.}\ \bibnamefont
  {Yang}}, \bibinfo {author} {\bibfnamefont {T.~M.}\ \bibnamefont {Rice}}, \
  and\ \bibinfo {author} {\bibfnamefont {F.-C.}\ \bibnamefont {Zhang}},\ }\href
  {\doibase 10.1103/PhysRevB.73.174501} {\bibfield  {journal} {\bibinfo
  {journal} {Phys. Rev. B}\ }\textbf {\bibinfo {volume} {73}},\ \bibinfo
  {pages} {174501} (\bibinfo {year} {2006})}\BibitemShut {NoStop}%
\bibitem [{\citenamefont {Rice}\ \emph {et~al.}(2012)\citenamefont {Rice},
  \citenamefont {Yang},\ and\ \citenamefont {Zhang}}]{Rice2012}%
  \BibitemOpen
  \bibfield  {author} {\bibinfo {author} {\bibfnamefont {T.~M.}\ \bibnamefont
  {Rice}}, \bibinfo {author} {\bibfnamefont {K.-Y.}\ \bibnamefont {Yang}}, \
  and\ \bibinfo {author} {\bibfnamefont {F.-C.}\ \bibnamefont {Zhang}},\ }\href
  {https://doi.org/10.1088/0034-4885/75/1/016502} {\bibfield  {journal}
  {\bibinfo  {journal} {Reports on Progress in Physics}\ }\textbf {\bibinfo
  {volume} {75}},\ \bibinfo {pages} {016502} (\bibinfo {year}
  {2012})}\BibitemShut {NoStop}%
\bibitem [{\citenamefont {Dzyaloshinskii}(2003)}]{Dzyaloshinskii2003}%
  \BibitemOpen
  \bibfield  {author} {\bibinfo {author} {\bibfnamefont {I.}~\bibnamefont
  {Dzyaloshinskii}},\ }\href {\doibase 10.1103/PhysRevB.68.085113} {\bibfield
  {journal} {\bibinfo  {journal} {Phys. Rev. B}\ }\textbf {\bibinfo {volume}
  {68}},\ \bibinfo {pages} {085113} (\bibinfo {year} {2003})}\BibitemShut
  {NoStop}%
\bibitem [{\citenamefont {Konik}\ \emph {et~al.}(2006)\citenamefont {Konik},
  \citenamefont {Rice},\ and\ \citenamefont {Tsvelik}}]{Konik2006}%
  \BibitemOpen
  \bibfield  {author} {\bibinfo {author} {\bibfnamefont {R.~M.}\ \bibnamefont
  {Konik}}, \bibinfo {author} {\bibfnamefont {T.~M.}\ \bibnamefont {Rice}}, \
  and\ \bibinfo {author} {\bibfnamefont {A.~M.}\ \bibnamefont {Tsvelik}},\
  }\href {\doibase 10.1103/PhysRevLett.96.086407} {\bibfield  {journal}
  {\bibinfo  {journal} {Phys. Rev. Lett.}\ }\textbf {\bibinfo {volume} {96}},\
  \bibinfo {pages} {086407} (\bibinfo {year} {2006})}\BibitemShut {NoStop}%
\bibitem [{\citenamefont {Stanescu}\ \emph {et~al.}(2007)\citenamefont
  {Stanescu}, \citenamefont {Phillips},\ and\ \citenamefont
  {Choy}}]{Stanescu2007}%
  \BibitemOpen
  \bibfield  {author} {\bibinfo {author} {\bibfnamefont {T.~D.}\ \bibnamefont
  {Stanescu}}, \bibinfo {author} {\bibfnamefont {P.}~\bibnamefont {Phillips}},
  \ and\ \bibinfo {author} {\bibfnamefont {T.-P.}\ \bibnamefont {Choy}},\
  }\href {\doibase 10.1103/PhysRevB.75.104503} {\bibfield  {journal} {\bibinfo
  {journal} {Phys. Rev. B}\ }\textbf {\bibinfo {volume} {75}},\ \bibinfo
  {pages} {104503} (\bibinfo {year} {2007})}\BibitemShut {NoStop}%
\bibitem [{\citenamefont {Dave}\ \emph {et~al.}(2013)\citenamefont {Dave},
  \citenamefont {Phillips},\ and\ \citenamefont {Kane}}]{Dave2013}%
  \BibitemOpen
  \bibfield  {author} {\bibinfo {author} {\bibfnamefont {K.~B.}\ \bibnamefont
  {Dave}}, \bibinfo {author} {\bibfnamefont {P.~W.}\ \bibnamefont {Phillips}},
  \ and\ \bibinfo {author} {\bibfnamefont {C.~L.}\ \bibnamefont {Kane}},\
  }\href {\doibase 10.1103/PhysRevLett.110.090403} {\bibfield  {journal}
  {\bibinfo  {journal} {Phys. Rev. Lett.}\ }\textbf {\bibinfo {volume} {110}},\
  \bibinfo {pages} {090403} (\bibinfo {year} {2013})}\BibitemShut {NoStop}%
\bibitem [{\citenamefont {Huang}\ \emph {et~al.}(2022)\citenamefont {Huang},
  \citenamefont {Nave},\ and\ \citenamefont {Phillips}}]{Huang2022}%
  \BibitemOpen
  \bibfield  {author} {\bibinfo {author} {\bibfnamefont {E.~W.}\ \bibnamefont
  {Huang}}, \bibinfo {author} {\bibfnamefont {G.~L.}\ \bibnamefont {Nave}}, \
  and\ \bibinfo {author} {\bibfnamefont {P.~W.}\ \bibnamefont {Phillips}},\
  }\href {\doibase 10.1038/s41567-022-01529-8} {\bibfield  {journal} {\bibinfo
  {journal} {Nature Physics}\ }\textbf {\bibinfo {volume} {18}},\ \bibinfo
  {pages} {511} (\bibinfo {year} {2022})}\BibitemShut {NoStop}%
\bibitem [{\citenamefont {Setty}(2020)}]{Setty2020}%
  \BibitemOpen
  \bibfield  {author} {\bibinfo {author} {\bibfnamefont {C.}~\bibnamefont
  {Setty}},\ }\href {\doibase 10.1103/PhysRevB.101.184506} {\bibfield
  {journal} {\bibinfo  {journal} {Phys. Rev. B}\ }\textbf {\bibinfo {volume}
  {101}},\ \bibinfo {pages} {184506} (\bibinfo {year} {2020})}\BibitemShut
  {NoStop}%
\bibitem [{\citenamefont {Setty}(2021)}]{Setty2021}%
  \BibitemOpen
  \bibfield  {author} {\bibinfo {author} {\bibfnamefont {C.}~\bibnamefont
  {Setty}},\ }\href {\doibase 10.1103/PhysRevB.103.014501} {\bibfield
  {journal} {\bibinfo  {journal} {Phys. Rev. B}\ }\textbf {\bibinfo {volume}
  {103}},\ \bibinfo {pages} {014501} (\bibinfo {year} {2021})}\BibitemShut
  {NoStop}%
\bibitem [{\citenamefont {Yang}(2021)}]{Yang2021}%
  \BibitemOpen
  \bibfield  {author} {\bibinfo {author} {\bibfnamefont {K.}~\bibnamefont
  {Yang}},\ }\href {\doibase 10.1103/PhysRevB.103.024529} {\bibfield  {journal}
  {\bibinfo  {journal} {Phys. Rev. B}\ }\textbf {\bibinfo {volume} {103}},\
  \bibinfo {pages} {024529} (\bibinfo {year} {2021})}\BibitemShut {NoStop}%
\bibitem [{\citenamefont {Zhao}\ \emph {et~al.}(2022)\citenamefont {Zhao},
  \citenamefont {Yeo}, \citenamefont {Huang},\ and\ \citenamefont
  {Phillips}}]{Zhao2022}%
  \BibitemOpen
  \bibfield  {author} {\bibinfo {author} {\bibfnamefont {J.}~\bibnamefont
  {Zhao}}, \bibinfo {author} {\bibfnamefont {L.}~\bibnamefont {Yeo}}, \bibinfo
  {author} {\bibfnamefont {E.~W.}\ \bibnamefont {Huang}}, \ and\ \bibinfo
  {author} {\bibfnamefont {P.~W.}\ \bibnamefont {Phillips}},\ }\href {\doibase
  10.1103/PhysRevB.105.184509} {\bibfield  {journal} {\bibinfo  {journal}
  {Phys. Rev. B}\ }\textbf {\bibinfo {volume} {105}},\ \bibinfo {pages}
  {184509} (\bibinfo {year} {2022})}\BibitemShut {NoStop}%
\bibitem [{foo()}]{footnote}%
  \BibitemOpen
  \href@noop {} {\bibinfo  {journal} {Note that the two Fermi surfaces are
  obtained only from the empty or the completely full limit. We thank P. W.
  Phillips for his useful comments on this issue. And from Eq.(\ref{eq2}), we
  see the zeros of the Green's function give the Luttinger surface as
  $\xi_{\mathbf{k}}+\left(1-\left<n_{\mathbf{k},\bar{\sigma}}\right>\right)U=0$}\
  }\BibitemShut {NoStop}%
\bibitem [{app()}]{appendix}%
  \BibitemOpen
\bibfield  {journal} {  }\href@noop {} {\bibinfo  {journal} {See the
  supplementary materials for details}\ }\BibitemShut {NoStop}%
\bibitem [{\citenamefont {Zhou}\ \emph {et~al.}(2017)\citenamefont {Zhou},
  \citenamefont {Kanoda},\ and\ \citenamefont {Ng}}]{RMP2017}%
  \BibitemOpen
\bibfield  {journal} {  }\bibfield  {author} {\bibinfo {author} {\bibfnamefont
  {Y.}~\bibnamefont {Zhou}}, \bibinfo {author} {\bibfnamefont {K.}~\bibnamefont
  {Kanoda}}, \ and\ \bibinfo {author} {\bibfnamefont {T.-K.}\ \bibnamefont
  {Ng}},\ }\href {\doibase 10.1103/RevModPhys.89.025003} {\bibfield  {journal}
  {\bibinfo  {journal} {Rev. Mod. Phys.}\ }\textbf {\bibinfo {volume} {89}},\
  \bibinfo {pages} {025003} (\bibinfo {year} {2017})}\BibitemShut {NoStop}%
\bibitem [{\citenamefont {Greiner}\ \emph {et~al.}(2012)\citenamefont
  {Greiner}, \citenamefont {Neise},\ and\ \citenamefont
  {St{\"o}cker}}]{Greiner2012}%
  \BibitemOpen
  \bibfield  {author} {\bibinfo {author} {\bibfnamefont {W.}~\bibnamefont
  {Greiner}}, \bibinfo {author} {\bibfnamefont {L.}~\bibnamefont {Neise}}, \
  and\ \bibinfo {author} {\bibfnamefont {H.}~\bibnamefont {St{\"o}cker}},\
  }\href@noop {} {\emph {\bibinfo {title} {Thermodynamics and statistical
  mechanics}}}\ (\bibinfo  {publisher} {Springer Science \& Business Media},\
  \bibinfo {year} {2012})\BibitemShut {NoStop}%
\bibitem [{\citenamefont {Cooper}(1956)}]{Cooper1956}%
  \BibitemOpen
  \bibfield  {author} {\bibinfo {author} {\bibfnamefont {L.~N.}\ \bibnamefont
  {Cooper}},\ }\href {\doibase 10.1103/PhysRev.104.1189} {\bibfield  {journal}
  {\bibinfo  {journal} {Phys. Rev.}\ }\textbf {\bibinfo {volume} {104}},\
  \bibinfo {pages} {1189} (\bibinfo {year} {1956})}\BibitemShut {NoStop}%
\end{thebibliography}%

\clearpage
\section{--Supplemental Materials--}

\subsection{Appendix A. The Green's function of the HK model}
\setcounter{equation}{0}
\setcounter{figure}{0}
\renewcommand {\theequation} {A.\arabic{equation}}
\renewcommand {\thefigure} {S\arabic{figure}}

Consider the the single-particle Green's function in the imaginary time space,%
\begin{equation}
G_{\sigma}\left(  \mathbf{k},\tau\right)  =-\Theta\left(  \tau\right)
\left\langle \hat{T}\left\{  c_{\mathbf{k}\sigma}\left(  \tau\right)
,c_{\mathbf{k}\sigma}^{\dagger}\left(  0\right)  \right\}  \right\rangle .
\end{equation}
where $\hat{T}$ is the time-ordering operator. By use of the equation of
motion,
\begin{align}
\partial_{\tau}G_{\sigma}\left(  \mathbf{k},\tau\right)   &  =-\delta\left(
\tau\right)  -\Theta\left(  \tau\right)  \left\langle \hat{T}\left\{
\partial_{\tau}c_{\mathbf{k}\sigma}\left(  \tau\right)  ,c_{\mathbf{k}\sigma
}^{\dagger}\left(  0\right)  \right\}  \right\rangle \nonumber\\
&  =-\delta\left(  \tau\right)  -\xi_{\mathbf{k}}G_{\sigma}\left(
\mathbf{k},\tau\right)  -UQ_{\sigma}\left(  \mathbf{k},\tau\right)  ,
\end{align}
where $Q_{\sigma}\left(  \mathbf{k},\tau\right)  $ is defined as%
\begin{equation}
Q_{\sigma}\left(  \mathbf{k},\tau\right)  \equiv-\Theta\left(  \tau\right)
\left\langle \hat{T}\left\{  n_{\mathbf{k}\bar{\sigma}}\left(  \tau\right)
c_{\mathbf{k}\sigma}\left(  \tau\right)  ,c_{\mathbf{k}\sigma}^{\dagger
}\left(  0\right)  \right\}  \right\rangle .
\end{equation}
Then, consider the equation of motion on $Q_{\sigma}\left(  \mathbf{k},\tau\right)
$, it gives%
\begin{equation}
\partial_{\tau}Q_{\sigma}\left(  \mathbf{k},\tau\right)=-\delta\left(  \tau\right)  \left\langle n_{\mathbf{k}\bar{\sigma}%
}\right\rangle -\left(  \xi_{\mathbf{k}}+U\right)  Q_{\sigma}\left(
\mathbf{k},\tau\right) .
\end{equation}

Perform the Fourier transformation by using the relation $Q_{\sigma}\left(  \mathbf{k}%
,\tau\right)  =\frac{1}{\beta}\sum_{\omega_{n}}e^{-i\omega_{n}\tau}Q_{\sigma
}\left(  \mathbf{k},i\omega_{n}\right)  $, where $\omega_n=(2n+1)\pi T$ ($n\in \mathbb{Z}$) is the fermionic Matsubara frequency and $T$ is the temperature, one can
obtain the analytical expression on $Q_{\sigma}\left(  \mathbf{k},i\omega
_{n}\right)  $,%
\begin{equation}
Q_{\sigma}\left(  \mathbf{k},i\omega_{n}\right)  =\frac{\left\langle
n_{\mathbf{k}\bar{\sigma}}\right\rangle }{i\omega_{n}-\xi_{\mathbf{k}}-U},
\end{equation}
and substitute it into the equation of motion of $G_{\sigma}$, one can obtain
the Green's function%
\begin{equation}
G_{\sigma}\left(  \mathbf{k},i\omega_{n}\right)  =\frac{1-\left\langle
n_{\mathbf{k}\bar{\sigma}}\right\rangle }{i\omega_{n}-\xi_{\mathbf{k}}}%
+\frac{\left\langle n_{\mathbf{k}\bar{\sigma}}\right\rangle }{i\omega_{n}%
-\xi_{\mathbf{k}}-U}.
\end{equation}
The ground state of the HK model can be written as%
\begin{equation}
\left\vert g\right\rangle =\left(  \prod_{\mathbf{k}_{2}\epsilon\Omega_{2}%
}c_{\mathbf{k}_{2}\uparrow}^{\dagger}c_{\mathbf{k}_{2}\downarrow}^{\dagger
}\right)  \left[  \prod_{\mathbf{k}_{1}\epsilon\Omega_{1}}\frac{1}{\sqrt{2}%
}\left(  c_{\mathbf{k}_{1}\uparrow}^{\dagger}+e^{i\phi_{\mathbf{k}}%
}c_{\mathbf{k}_{1}\downarrow}^{\dagger}\right)  \right]  \left\vert
0\right\rangle ,
\end{equation}
where $\phi_{\mathbf{k}}$ is an\ arbitrary phase which has no effect on the
physical quantities and we take $\phi_{\mathbf{k}}=0$ for brevity, and
$\left\vert 0\right\rangle $ is the vacuum state. Due the spin uncertainty in  the $\Omega_1$ region, there are large spin degeneracies are persist in the low temperature.

Since $H_{\text{HK}}\left(  \mathbf{k}\right)  $ are composed of 4 states for each $\mathbf{k}$ point: $|0\rangle,c^{\dagger}_{\bf{k},\uparrow}|0\rangle,c^{\dagger}_{\bf{k},\downarrow}|0\rangle,c^{\dagger}_{\bf{k},\uparrow}c^{\dagger}_{\bf{k},\downarrow}|0\rangle$, with their energies are $0$, $\xi_{\mathbf{k}}$, $\xi_{\mathbf{k}}$ and $2\xi_{\mathbf{k}}+U$, respectively. One can easily obtain the partition function  as%
\begin{equation}
Z_{\text{HK}}=\text{Tr}\left(  e^{-\beta H_{\text{HK}}}\right)  =
{\textstyle\prod\limits_{\mathbf{k}}}
Z_{\text{HK},\mathbf{k}},
\end{equation}
with
\begin{equation}
Z_{\text{HK},\mathbf{k}}=\text{Tr}\left[  e^{-\beta H_{\text{HK}}\left(  \mathbf{k}%
\right)  }\right]  =1+2e^{-\beta\xi_{\mathbf{k}}}+e^{-\beta\left(
2\xi_{\mathbf{k}}+U\right)  }.
\end{equation}
where $\beta=1/k_{\text{B}}T$, with $k_{\text{B}}$ is  the
Boltzmann constant, which is taken as $1$ for brevity.

The particle density distribution can be expressed as%
\begin{align}
\left\langle n_{\mathbf{k}\sigma}\left(  T\right)  \right\rangle &=\frac{1}%
{Z_{\text{HK}}}\text{Tr}\left(  n_{\mathbf{k}\sigma}e^{-\beta H_{\text{HK}}}\right)\nonumber\\
&=\frac{e^{-\beta\xi_{\mathbf{k}}}+e^{-\beta\left(  2\xi_{\mathbf{k}}+U\right)  }} {1+2e^{-\beta\xi_{\mathbf{k}}} +e^{-\beta\left(2\xi_{\mathbf{k}}+U\right)}}.
\end{align}%

In the zero-temperature limit, for repulsive
interaction $\left(  U>0\right)  $,%
\begin{align}
\left\langle n_{\mathbf{k}\sigma}\right\rangle
&  =\frac{1}{2}\left[  \Theta\left(  -\xi_{\mathbf{k}}\right)  +\Theta\left(
-\xi_{\mathbf{k}}-U\right)  \right] \nonumber\\
& =\left\{
\begin{array}
[c]{cc}%
0, & \xi_{\mathbf{k}}>0\\
\frac{1}{2}, & -U<\xi_{\mathbf{k}}<0\\
1 & \xi_{\mathbf{k}}<-U
\end{array}
\right.  ,
\end{align}

\subsection{Appendix B: Revisit the Cooper instability}
\setcounter{equation}{0}
\renewcommand {\theequation} {B.\arabic{equation}}
Before the study of the superconductivity, we investigate the microscopic picture of Cooper instability from the unconventional metal of the HK model.

With adding a BCS pairing interaction $H_{\text{pairing}}=-V\sum_{\mathbf{k},\mathbf{k}^{\prime}}c^{\dagger}_{\mathbf{k}\uparrow}c^{\dagger}_{-\mathbf{k}\downarrow}c_{-\mathbf{k}^{\prime}\downarrow}c_{\mathbf{k}^{\prime}\uparrow}$, The average energy of HK-BCS
model can be evaluated as%
\begin{equation}
E_{0}=\left\langle g\right\vert H_{\text{HK}}+H_{\text{pairing}}\left\vert
g\right\rangle =\sum_{\mathbf{k}\epsilon\Omega_{1}}\xi_{\mathbf{k}}%
+\sum_{\mathbf{k}\epsilon\Omega_{2}}\left(  2\xi_{\mathbf{k}}+U\right)
-n_{0}V.
\end{equation}
where $n_{0}=\frac{1}{4}\sum_{\mathbf{k}\epsilon\Omega_{1}}+\sum
_{\mathbf{k}\epsilon\Omega_{2}}$.
Along with the Cooper's approach\cite{Cooper1956,Phillips2020}, we construct the Cooper-pair wavefunction,%
\begin{equation}
\left\vert \psi\right\rangle =\sum_{\mathbf{k}\epsilon\Omega_{0}}%
\alpha_{\mathbf{k}}b_{\mathbf{k}}^{\dagger}\left\vert g\right\rangle
+\sum_{\mathbf{k}\epsilon\Omega_{1}}\beta_{\mathbf{k}}b_{\mathbf{k}}^{\dagger
}\left\vert g\right\rangle,
\end{equation}
where $b_{\mathbf{k}}^{\dagger}=c^{\dagger}_{\mathbf{k}\uparrow}c^{\dagger}_{\mathbf{k}\downarrow}$, and the normalization condition for $\left\vert \psi\right\rangle $ gives the relations for the coefficients $\alpha_{\mathbf{k}}$, $\beta_{\mathbf{k}}$,%
\begin{equation}
\left\langle \psi|\psi\right\rangle =\sum_{\mathbf{k}\epsilon\Omega_{0}%
}\left\vert \alpha_{\mathbf{k}}\right\vert ^{2}+\frac{1}{4}\sum_{\mathbf{k}%
\epsilon\Omega_{1}}\left\vert \beta_{\mathbf{k}}\right\vert ^{2}=1.
\end{equation}

The average energy of the $\left\vert \psi\right\rangle $ can be evaluated as
\begin{align}
E  =&\left\langle \psi\right\vert H\left\vert \psi\right\rangle \nonumber\\
=&\sum_{\mathbf{k}\epsilon\Omega_{0}} 2\xi_{\mathbf{k}}
\left\vert \alpha_{\mathbf{k}}\right\vert ^{2}+\frac{1}{4}\sum_{\mathbf{k}%
\epsilon\Omega_{1}}\left(  \xi_{\mathbf{k}}+U\right)  \left\vert
\beta_{\mathbf{k}}\right\vert ^{2}\nonumber\\
&-V\sum_{\mathbf{k,k}^{\prime}\epsilon
\Omega_{0}}\alpha_{\mathbf{k}}^{\ast}\alpha_{\mathbf{k}^{\prime}}-\frac{V}%
{16}\sum_{\mathbf{k,k}^{\prime}\epsilon\Omega_{0}}\beta_{\mathbf{k}}^{\ast
}\beta_{\mathbf{k}^{\prime}} \nonumber\\
&  -\frac{V}{4}\sum_{\mathbf{k}\epsilon\Omega_{0},\mathbf{k}^{\prime}%
\epsilon\Omega_{1}}\left(  \alpha_{\mathbf{k}}^{\ast}\beta_{\mathbf{k}%
^{\prime}}+\alpha_{\mathbf{k}}\beta_{\mathbf{k}^{\prime}}^{\ast}\right)
+E_{0}.
\end{align}
Then, the energy change with adding two electrons is $E_{\text{C}}=E-E_{0}$.
Including the normalization condition by introduce a Lagrange multipler
$\lambda$, we define the function
\begin{equation}
Q  =E_{\text{C}}-\lambda\left(  \left\langle \psi|\psi\right\rangle -1\right),
\end{equation}
and using the variational conditions $\frac{\partial Q}{\partial\alpha_{\mathbf{k}}^{\ast}}=0$ and $\frac{\partial Q}{\partial\beta_{\mathbf{k}}^{\ast}}=0$, the equations for $\alpha_{\mathbf{k}}$ and $\beta_{\mathbf{k}}$ can be obtained as
\begin{equation}
\alpha_{\mathbf{k}}=\frac{V}{2\xi_{\mathbf{k}}-E_{\text{C}}}\left(  \sum
_{\mathbf{k}^{\prime}\epsilon\Omega_{0}}\alpha_{\mathbf{k}^{\prime}}+\frac
{1}{4}\sum_{\mathbf{k}^{\prime}\epsilon\Omega_{1}}\beta_{\mathbf{k}^{\prime}%
}\right)  ,
\end{equation}%
\begin{equation}
\beta_{\mathbf{k}}=\frac{V}{\xi_{\mathbf{k}}+U-E_{\text{C}}}\left(  \sum
_{\mathbf{k}^{\prime}\epsilon\Omega_{0}}\alpha_{\mathbf{k}^{\prime}}+\frac
{1}{4}\sum_{\mathbf{k}^{\prime}\epsilon\Omega_{1}}\beta_{\mathbf{k}^{\prime}%
}\right)  .
\end{equation}
Summing $\alpha_{\mathbf{k}}$ and $\beta_{\mathbf{k}}$ over $\Omega_{0}$ and
$\Omega_{1}$ regions respectively, we can obtain the self-consistent equation%
\begin{equation}
1=\sum_{\mathbf{k}\epsilon\Omega_{0}}\frac{V}{2\xi_{\mathbf{k}}-E_{\text{C}}}+\frac{1}{4}\sum_{\mathbf{k}\epsilon\Omega_{1}}\frac{V}{\xi_{\mathbf{k}%
}+U-E_{\text{C}}}.
\label{eqA9}
\end{equation}

For brevity, if we consider $\rho\left(  \omega\right)  =\sum_{\mathbf{k}%
}\delta\left(  \omega-\varepsilon_{\mathbf{k}}\right)  =\frac{1}{W}$ for
$-W/2<\omega<W/2$, and assume $U<W$ and $W>2\mu$, then,

\begin{align}
\rho_{0}\left(  \omega\right)  =&\sum_{\mathbf{k}\in\Omega_{0}}\delta\left(\omega-\xi_{\mathbf{k}}\right)  =\Theta\left(  \omega\right) \rho\left(\omega+\mu\right),\\
\rho_{1}\left(  \omega\right)    =&\sum_{\mathbf{k}\in\Omega_{1}}\frac{1}%
{2}\left[  \delta\left(  \omega-\xi_{\mathbf{k}}\right)  +\delta\left(
\omega-\xi_{\mathbf{k}}-U\right)  \right] \nonumber\\
 =&\frac{1}{2}\left[  \Theta\left(  -\omega\right)  \Theta\left(
\omega+U\right)  \rho\left(  \omega+\mu\right) \right. \nonumber\\
&+\left.\Theta\left(  -\omega
+U\right)  \Theta\left(  \omega\right)  \rho\left(  \omega+\mu-U\right)
\right].
\end{align}

By using the relation $\sum_{\mathbf{k}\in\Omega_{i}}(...)=\int \rho_{i}(\omega)(...)d\omega$, the self consistent equation eq.(\ref{eqA9}) can then be rearranged as%
\begin{align}
1  =& \int\frac{V\rho_{0}\left(  \omega\right)  }{2\omega-2\mu-E_{\text{C}}}d\omega+\frac{1}{4}\int\frac{V\rho_{1}\left(  \omega\right)  }{\omega
-\mu+U-E_{\text{C}}}d\omega\nonumber\\
 =& \frac{V}{W}\int_{0}^{\frac{W}{2}-\mu}\frac{d\omega}{2\omega-E_{\text{C}}} +\frac{V}{8W}\int_{-U}^{0}\frac{d\omega}{\omega+U-E_{\text{C}}}\nonumber\\
& +\frac{V}{8W}%
\int_{0}^{U}\frac{d\omega}{\omega-E_{\text{C}}}\nonumber\\
 =& \frac{V}{4W}\ln\left\vert \frac{\left(  W-2\mu-E_{\text{C}}\right)
^{2}\left(  U-E_{\text{C}}\right)  }{E_{\text{C}} ^{3}}\right\vert .
\end{align}

Define the dimensionless quantities $u=U/W$,
$\upsilon=V/W$, $\varepsilon=E_{\text{C}}/W$, $\tilde{\mu}=\mu/W$, the above equation reduces as%
\begin{equation}
1=\frac{\upsilon}{4}\ln\left\vert \frac{\left(  1-2\tilde{\mu}-\varepsilon\right)
^{2}\left(  u-\varepsilon\right)  }{\varepsilon^{3}}\right\vert .
\end{equation}

In BCS case, $u=0$, the equation gives $1=\frac{\upsilon}{2}\ln\left\vert \frac{1-2\tilde{\mu}-\varepsilon}{-\varepsilon
}\right\vert$, which yielding the solution: $\varepsilon=-\frac{1-2\tilde{\mu}}{e^{\frac{2}{\upsilon}+1}}\approx-\left(1-2\tilde{\mu}\right)e^{-\frac{2}{\upsilon}%
}$. One can see that, an infinitesimal attraction can induce a stable Cooper-pair
bound state $\left(  \text{i.e., }\varepsilon<0\right)  $.

For $u\neq0$, as $\upsilon<<1$, and $\left\vert \varepsilon\right\vert <<u$,
the negative solution for $\varepsilon$ is given by the equation $-e^{-\frac{4}{\upsilon}}\approx\frac{\varepsilon^{3}}{\left(  1-2\tilde{\mu}\right)^{2}u}$, which yielding to
$\varepsilon=-\left( 1-2\tilde{\mu}\right)  ^{2/3}u^{1/3}e^{-\frac{4}{3\upsilon}}$. As is shown in here and in main text, the Cooper instability is still exact for finite $U$ case.

\subsection{Appendix C: Exact diagonalization of the mean-field HK-BCS model}
\setcounter{equation}{0}
\renewcommand {\theequation} {C.\arabic{equation}}

In the mean-field level, the HK-BCS model can be rearranged as the summation over one-half of the Brillouin zone,
\begin{align}
H=\sum_{\bf{k}\in\frac{1}{2}\text{BZ}}&\left[\xi_{\bf{k}}\left(n_{\bf{k},\uparrow}+n_{\bf{-k},\uparrow}+n_{\bf{k},\downarrow} +n_{\bf{-k},\downarrow}\right)\right.\nonumber\\
& +U\left(n_{\bf{k}\uparrow}n_{\bf{k}\downarrow} +n_{\bf{-k}\uparrow}n_{\bf{-k}\downarrow} \right)\nonumber\\
&\left.+\left(\Delta c^{\dagger}_{\bf{k}\uparrow} c^{\dagger}_{-\bf{k}\downarrow}+\Delta c^{\dagger}_{\bf{-k}\uparrow} c^{\dagger}_{\bf{k}\downarrow}+\text{H.c.}\right)\right],
\end{align}
which can be exactly diagonalized in the Fock space spanned by the 4-fermion occupation states $\{\left\vert n_{\bf{k},\uparrow},n_{\bf{k},\downarrow},n_{\bf{-k},\uparrow},n_{\bf{-k},\downarrow}\right>\}$ for each $\bf{k}$ point.

By performing the exact diagonalization, we can obtain the eigen states $\{\left|n,\mathbf{k}\right>\}$ and the eigenspectra $\{E_{n,\mathbf{k}}\}$. Then, the free energy can be computed as,
\begin{equation}
    F_{\rm S}\left[\Delta\right]=-T\ln Z=-T\sum_{\mathbf{k}\in\frac{1}{2}\text{BZ}}\ln Z_{\mathbf{k}},
\end{equation}
where $Z_{\mathbf{k}}=\sum_{n}e^{-E_{n,\mathbf{k}}/T}$. And the superconducting gap $\Delta\equiv-V\sum_{\mathbf{k}}\left<c_{-\mathbf{k}\downarrow}c_{\mathbf{k}\uparrow}\right>$ can be found by searching the global minimum of the free energy, with the help of its minimization,
\begin{equation}
\frac{\partial F_{\rm S}\left[\Delta\right]}{\partial \Delta}=0.
\end{equation}

\begin{figure}
\includegraphics[width=0.46\textwidth]{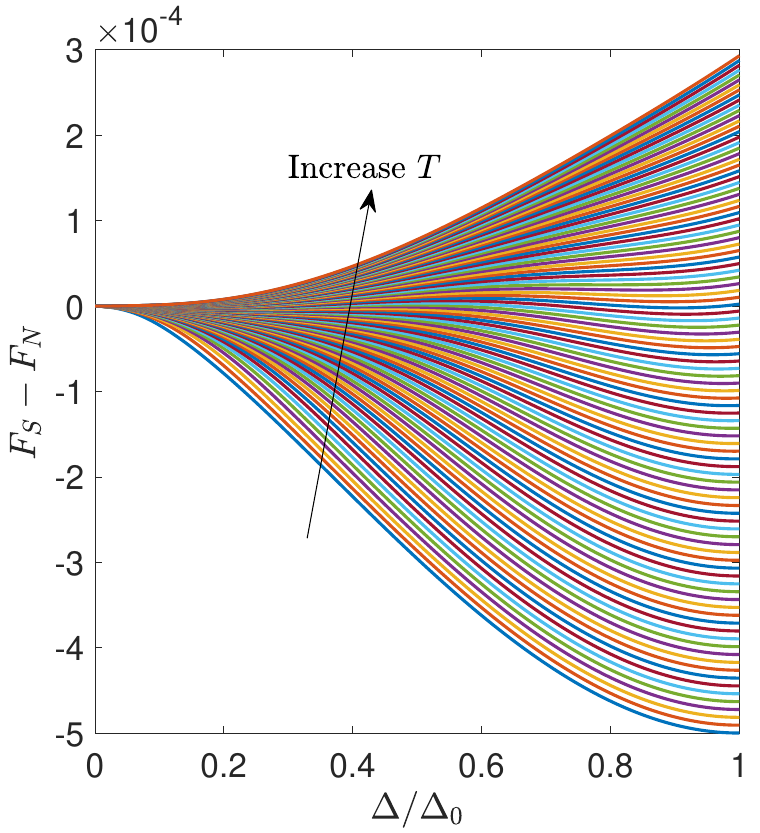}
\caption{The evolution of the changes of the free energy as a function of $\Delta/\Delta_0$ for $U=4$, $n=0.4$ and $V=1$ with different temperatures.}
\label{figS1}
\end{figure}

Fig.~\ref{figS1} show the numerical data of the evolution of the changes of the free energy $F=F_{\rm S}-F_{\rm N}$ for $U=4$, $n(T=0)=0.4$ and $V=1$, where the $F_{\rm N}$ refers to the normal-state free energy with taken $\Delta=0$.

\subsection{Appendix D: Ginzburg-Landau analysis}
\setcounter{equation}{0}
\renewcommand {\theequation} {D.\arabic{equation}}
Define the Ginzburg-Landau free energy functional as,
\begin{equation}
\delta{}\mathcal{F}\left[  \Delta\right]  =\alpha\Delta^{2}+\frac{\beta}{2}\Delta^{4}+\frac{\gamma}{3!}\Delta
^{6}+\frac{\eta}{4!}\Delta^{8}+O\left(  \Delta^{8}\right),
\end{equation}
where $\eta>0$, or $\eta=0$ and $\gamma>0$, ensures the stability of the system, and $\alpha$, $\beta$, $\gamma$ are the expansion coefficients.

For brevity, we introduce $x\equiv\Delta^2$, and define a function $f(x)$ as
\begin{equation}
f\left(x\right) =\alpha x+\frac{\beta}{2}x^{2}+\frac{\gamma}{3!}x^{3}+\frac{\eta}{4!}x^{4}.
\label{eqD2}
\end{equation}
Then, the minimization of the free energy is equivalent to find the minimum of $f\left(x\right)$ for $x\geq0$. By requiring $f^{\prime}\left(  x\right)=\frac{\partial f\left(  x\right)}{ \partial x}=0$, the extreme of  $f\left(x\right)$ is determined by the cubic equation,
\begin{equation}
f^{\prime}\left(  x\right)=\alpha+\beta x+\frac{\gamma}{2}x^{2}+\frac{\eta}{6}x^{3}=0.
\label{eqD3}
\end{equation}

\begin{figure}
\includegraphics[width=0.46\textwidth]{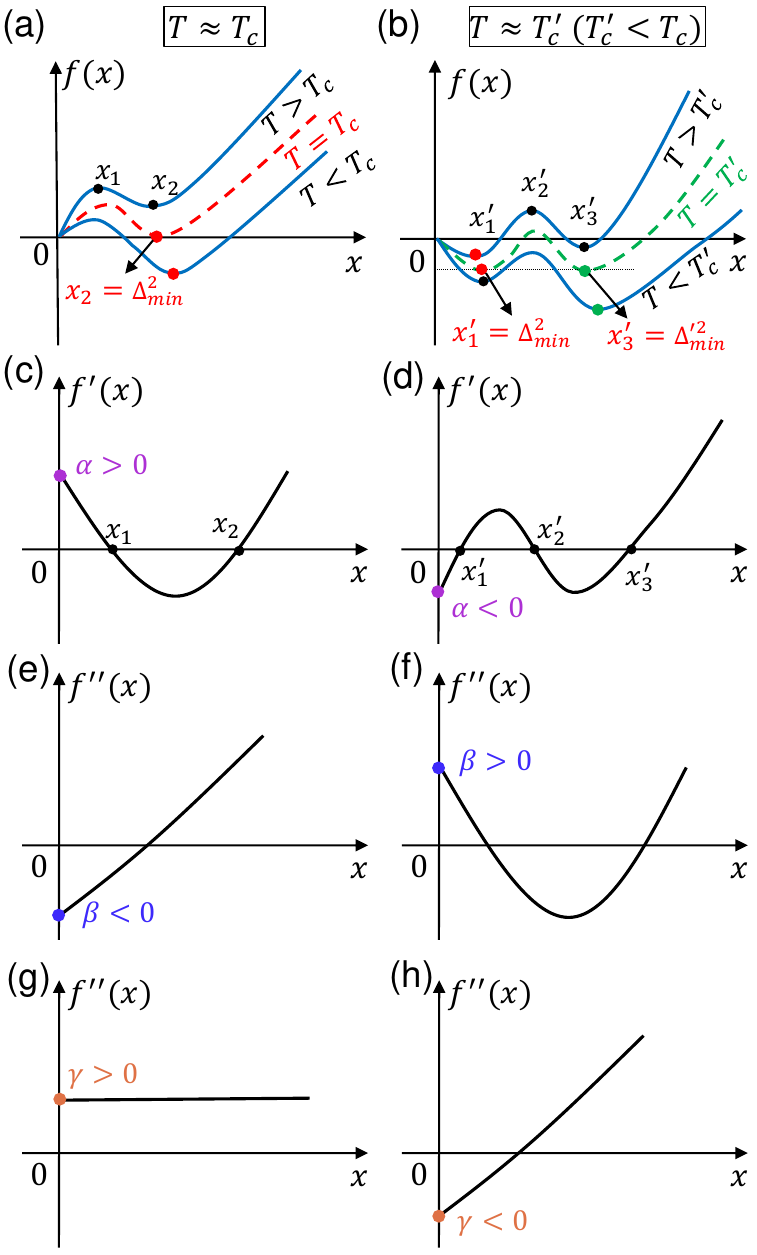}
\caption{The sketched function $f(x)$ [defined in Eq.~\eqref{eqD2} with $x=\Delta^2$] in two critical regions: (a) $T\approx{}T_{\rm c}$ and (b) $T\approx{}T_{\rm c}^{\prime}$ ($T_{\rm c}^{\prime}<T_{\rm c}$). At the second critical region, $T\approx{}T_{\rm c}^{\prime}$, there are two separated local minimums at $\Delta_{\rm min}$ and $\Delta_{\rm min}^{\prime}$ ($\Delta_{\rm min}<\Delta_{\rm min}^{\prime}$).(c) $f^{\prime}(x)$ at $T\approx{}T_{\rm c}$ and (d) $f^{\prime}(x)$ at $T\approx{}T_{\rm c}^{\prime}$. At $T\approx{}T_{\rm c}$,
(e) $\beta<0$ and (g) $\gamma>0$ are guaranteed only when $\eta=0$; while at $T\approx{}T_{\rm c}^{\prime}$, (f) $\beta>0$ and (h) $\gamma<0$ always hold, as well as $\eta>0$.}
\label{figS2}
\end{figure}
To study phase transitions for the HK-BCS model, we consider critical regions: $T\approx T_{\rm c}$ and $T\approx T_{\rm c}^{\prime}$.

(1) $\mathbf{T\approx T_{\rm c}}$: to make the first-order transition happen, the superconducting order parameter will be switched to a finite value $\Delta=\Delta_{\rm min}$ from $\Delta=0$. There are at least one maximal $x=x_1$ and one minimum $x=x_2$ occur ($x_2>x_1>0$), as shown in Fig.~\ref{figS2}(a), and the critical condition is determined by the solution of the equations $f(x_2)=f^{\prime}(x_2)=0$, i.e.,
\begin{align}
\alpha+\frac{\beta}{2}x_2+\frac{\gamma}{6}x_2^2+\frac{\eta}{24}x_2^3&=0,\\
\alpha+\beta x_2+\frac{\gamma}{2}x_2^2+\frac{\eta}{6}x_2^3&=0.
\end{align}
Combining the above two equations to eliminate $\alpha$ or $\eta$, we can obtain the two equations
\begin{align}
\frac{1}{2}\beta+\frac{\gamma}{3}x_2+\frac{\eta}{8}x_2^2&=0,\\
3\alpha+\beta x_2+\frac{\gamma}{6}x_2^2&=0.
\end{align}
These two quadratic equations should share a same positive root $x_2$, i.e.,
\begin{equation}
x_2=\frac{3\left(9\alpha\eta-2\gamma\beta\right)}{4\gamma^2-9\beta\eta}=\frac{12\left(\beta^2- 2\alpha\gamma\right)}{9\alpha\eta-2\beta\gamma}>0.
\label{eqD8}
\end{equation}

The critical point can be archived under the condition,
\begin{equation}
\left(9\alpha\eta-2\gamma\beta\right)^2=4\left(\beta^2- 2\alpha\gamma\right)\left(4\gamma^2-9\beta\eta\right),
\end{equation}
where $\alpha>0$ and $\eta\geq0$, and the superconducting gap at $T_{\rm c}$ reads
\begin{equation}
\Delta(T=T_{\rm c})=\sqrt{\frac{3\left(9\alpha\eta-2\gamma\beta\right)}{4\gamma^2-9\beta\eta}} =\sqrt{\frac{12\left(\beta^2- 2\alpha\gamma\right)}{9\alpha\eta-2\beta\gamma}}.
\end{equation}
In the limit of $\eta=0$, it becomes
\begin{equation}
\Delta(T=T_{\rm c})=\sqrt{\frac{-3\beta}{2\gamma}} =\sqrt{\frac{6\left(2\alpha\gamma-\beta^2 \right)}{\beta\gamma}},
\end{equation}
which restores the result in Ref.~\cite{Zhao2022}.

(2) $\mathbf{T\approx T_{\rm c}^{\prime}}$: in this case, the second first-order-like change takes place only when $f(x)$ displays two minimums ($x_1^{\prime}$, $x_3^{\prime}$) and one maximum ($x_2^{\prime}$) for $x>0$ ($x_3^{\prime}>x_2^{\prime}>x_1^{\prime}>0$), which requires $\eta>0$. As shown in Fig.~\ref{figS2}(b), and the superconducting gap is switched from $\Delta^{-}_{min}=\sqrt{x_1^{\prime}}$ to $\Delta^{+}_{min}=\sqrt{x_3^{\prime}}$ as decreasing $T$. Moreover, as shown in Figs.~\ref{figS2}(d), (f) and (h), and the sign of  $\alpha$, $\beta$ and $\gamma$ can be determined,
\begin{equation}
\alpha<0,~~\beta>0,~~\gamma<0.
\label{eqD9}
\end{equation}

At $T=T_{\rm c}^{\prime}$, consider the minima $f_0\equiv f(x_1^{\prime})=f(x_3^{\prime})<0$ and the fact $f^{\prime}(x_1^{\prime})=f^{\prime}(x_3^{\prime})=0$, we have
\begin{equation}
f(x)-f_0=\frac{\eta}{24}\left(x-x_1^{\prime}\right)^2\left(x-x_3^{\prime}\right)^2,
\label{eqD14}
\end{equation}
where from $f(0)=0$, $f_0=-\frac{\eta}{24}x_1^{\prime2}x_3^{\prime2}$ can be obtained.
Compare the coefficients in Eq.~(\ref{eqD2}) and Eq.~(\ref{eqD14}), we obtain the following relations,
\begin{align}
x_1^{\prime}x_3^{\prime}\left(x_1^{\prime}+x_3^{\prime}\right)&=-\frac{12\alpha}{\eta},\\
x_1^{\prime2}+x_3^{\prime2}+4x_1^{\prime}x_3^{\prime}&=-\frac{12\beta}{\eta},\\
x_1^{\prime}+x_3^{\prime}&=-\frac{2\gamma}{\eta}.
\end{align}
Thus, the constraint in Eq.~(\ref{eqD9}) can also be verified from the above three equations. The solution to $x_1^{\prime}$ and $x_3^{\prime}$ can be computed, i.e.,
\begin{align}
x_1^{\prime}&=-\frac{\gamma}{\eta}-\sqrt{\frac{\gamma^2}{\eta^2}-\frac{6\alpha}{\gamma}},\label{eqD18}\\
x_3^{\prime}&=-\frac{\gamma}{\eta}+\sqrt{\frac{\gamma^2}{\eta^2}-\frac{6\alpha}{\gamma}}\label{eqD19},
\end{align}
and the consistency constraint gives rise to the critical condition at $T=T_{\rm c}^{\prime}$,
\begin{equation}
\frac{\gamma^2}{\eta^2}=3\left(\frac{\beta}{\eta}-\frac{\alpha}{\gamma}\right).
\label{eqD20}
\end{equation}
Then around $T_{\rm c}^{\prime}$, the superconducting order parameter $\Delta=\Delta_{\rm min}(=\sqrt{x_1^{\prime}})$ for $T=T^{\prime +}_c$, and $\Delta=\Delta^{\prime}_{min}(=\sqrt{x_3^{\prime}})$ for $T= T^{\prime -}_c$.
It is worth noting that the first-order-like change at $T=T_{\rm c}^{\prime}$ will vanish and become a crossover when $\Delta_{\rm min}=\Delta_{\min}^{\prime}$, which leads to an additional condition for the crossover,
\begin{equation}
\gamma^3=6\alpha\eta^2.
\end{equation}

Then, we continue to discuss the critical region $T\approx{}T_{\rm c}^{\prime}$ by introducing a dimensionless parameter $t^{\prime}=\left(T-T_{\rm c}^{\prime}\right)/T_{\rm c}^{\prime}$. As the temperature goes up (or down) from $T=T_{\rm c}^{\prime}$, the global minimum become $x_1^{\prime}$ ($x_3^{\prime}$), and the line of $x_1^{\prime}\rightarrow x_3^{\prime}$ tilted up (down), similar to previous discussions for $T=T_{\rm c}$, by subtracting to the tangent line $x_1^{\prime}\rightarrow x_3^{\prime}$, the free energy can be approximately written in the form as,
\begin{equation}
f(x)-\epsilon t^{\prime}x=\frac{\eta}{24}\left(x-x_1^{\prime}\right)^2\left(x-x_3^{\prime}\right)^2+f_0,
\end{equation}
where
\begin{equation}
\epsilon=\left\{
                 \begin{array}{cc}
                   \epsilon_{+}, &\text{for }T>T_{\rm c}^{\prime}\\
                   \epsilon_{-}, &\text{for }T<T_{\rm c}^{\prime}\\
                 \end{array}
          \right.,
\end{equation}
in which $\epsilon_{\pm}>0$ and $\epsilon_{+}\neq \epsilon_{-}$. This approximate form will replace the coefficient $\alpha\rightarrow\alpha+\epsilon t^{\prime}$. Substitute it into the critical condition (\ref{eqD20}), we find that
\begin{equation}
\alpha=\gamma\left( \frac{\beta}{\eta}-\frac{\gamma^2}{3\eta^2} \right)-\epsilon t^{\prime}.
\end{equation}
When $T>T_{\rm c}^{\prime}$, to the leading order of $t^{\prime}$, $\Delta=\sqrt{x_1^{\prime}}$ can be approximated as
\begin{align}
\Delta\left(t^{\prime}\right)&=\sqrt{-\frac{\gamma}{\eta}-\sqrt{\frac{\gamma^2}{\eta^2}-\frac{6(\alpha+\epsilon_+ t^{\prime})}{\gamma}}}\nonumber\\
&\approx \Delta_{\rm min}\left[ 1+ \frac{\eta^2\epsilon_{+}}{2\Delta_{\rm min}^{2}\gamma\left(\gamma^2-2\beta\eta\right)}t^{\prime}\right],
\end{align}
and when $T<T_{\rm c}^{\prime}$,
\begin{equation}
\Delta\left(t^{\prime}\right)\approx \Delta^{\prime}_{min}\left[ 1+ \frac{\eta^2\epsilon_{-}}{2\Delta_{\rm min}^{\prime{}2}\gamma\left(\gamma^2-2\beta\eta\right)}t^{\prime}\right].
\end{equation}

\end{document}